\documentclass[11pt,letterpaper,english]{article}
\usepackage[letterpaper,margin=1in]{geometry}
\usepackage[utf8]{inputenc}

\usepackage{hyperref}
\usepackage{graphicx}
\usepackage{amsmath}
\usepackage{amsfonts}
\usepackage{amsthm}
\usepackage{dsfont}
\usepackage{mathtools}
\usepackage{booktabs}
\usepackage{todonotes}
\usepackage{paralist}
\usepackage[ruled]{algorithm2e}

\theoremstyle{definition}

\theoremstyle{plain}
\newtheorem{assumption}{Assumption}
\newtheorem{lemma}{Lemma}
\newtheorem{theorem}{Theorem}

\newtheorem*{theorem*}{Theorem}
\newtheorem{corollary}{Corollary}
\theoremstyle{remark}

\renewcommand{\vec}[1]{\boldsymbol{\mathbf{#1}}}

\newcommand{\N}{\mathbb{N}}
\newcommand{\R}{\mathbb{R}}
\newcommand{\diag}{\mathop{\mathrm{diag}}}

\newcommand{\fwprec}{\epsilon}
\newcommand{\region}[1][\vec{t}]{R_{#1}}

\makeatletter
\newcommand{\mcflow}[1][e]{\@ifstar{x^*}{x^*_{#1}}}
\newcommand{\mcflowV}{\vec{x}^*}
\newcommand{\ecost}[1][e]{\@ifstar{F}{F_{#1}}}
\newcommand{\mcost}[1][e]{\@ifstar{f}{f_{#1}}}
\makeatother

\usepackage{tikz}
\usetikzlibrary{shapes.geometric, arrows, positioning, angles, quotes}
\usetikzlibrary{patterns}
\usetikzlibrary{decorations.pathmorphing}
\usetikzlibrary{fit,calc}
\usetikzlibrary{snakes}

\tikzset{
arrow/.style={-latex},
>=latex
}

\tikzstyle{vertex} = [circle, draw=black, thick, minimum height=1.5em, text centered]
\tikzstyle{solid} = =[circle, fill,inner sep=1.5pt,outer sep=0pt]

\title{Approximate Parametric Computation of Minimum-Cost Flows with~Convex Costs\footnote{Funded by the Deutsche Forschungsgemeinschaft (DFG, German Research
Foundation) under Germany's Excellence Strategy – The Berlin Mathematics
Research Center MATH+ (EXC-2046/1, project ID: 390685689).}}

\author{
    Per Joachims\footnote{Technische Universit\"at Berlin, Berlin, Germany, \texttt{p.joachims@campus.tu-berlin.de}}
    \and
	Max Klimm\footnote{Technische Universit\"at Berlin, Berlin, Germany, \texttt{$\{$klimm,philipp.warode$\}$@tu-berlin.de}}
	\and
	Philipp Warode\footnotemark[3]
}

\date{}

\begin{document}

\thispagestyle{plain}
\maketitle

\begin{abstract}
This paper studies a variant of the minimum-cost flow problem in a graph with convex cost function where the demands at the vertices are functions depending on a one-dimensional parameter~$\lambda$. We devise two algorithmic approaches for the approximate computation of parametric solutions for this problem. The first approach transforms an instance of the parametric problem into an instance with piecewise quadratic cost functions by interpolating the marginal cost functions. The new instance can be solved exactly with an algorithm we developed in prior work. In the second approach, we compute a fixed number of non-parametric solutions and interpolate the resulting flows yielding an approximate solution for the original, parametric problem. For both methods we formulate explicit bounds on the step sizes used in the respective interpolations that guarantee relative and absolute error margins. 
Finally, we test our approaches on real-world traffic and gas instances in an empirical study.
\end{abstract}


\section{Introduction}

This paper studies the approximate parametric computation of minimum-cost flows with separable convex costs as a function of the flow demand. We are given a strongly connected directed graph $G = (V,E)$ and for every edge~$e \in E$ a strictly convex cost function $F_e : \R_{\geq 0} \to \R_{\geq 0}$.
Let $\vec b^0 = (b_v^0)_{v \in V}$ and $\vec b = (b_v)_{v \in V}$ be two balanced demand vectors, i.e., $\sum_{v \in V} b^0_v = \sum_{v \in V} b_v = 0$. We are interested in computing a function $\vec x : \R_{\geq 0} \to \R_{\geq 0}$ such that for all $\lambda \in [0,1]$ the flow $\vec x(\lambda) = (x_e(\lambda))_{e \in E}$ is a minimum-cost flow for the demand vector $\vec b^0 + \lambda \vec b$, i.e., $\vec x(\lambda)$ solves the optimization problem
\begin{align}
\begin{array}{rrll}
\min & \displaystyle\sum_{e \in E} \ecost (x_e(\lambda)) \qquad\qquad\quad\;\,\\
\text{s.t.} & \displaystyle\sum_{e \in \delta^- (v)} x_e(\lambda) - \displaystyle\sum_{e \in \delta^+ (v)} x_e(\lambda) &= b^0_v + \lambda b_v
	&\quad\text{for all $v \in V$}, \\
& x_e(\lambda) &\geq 0 & \quad\text{for all $e \in E$}.
\end{array}\label{eq:mincost_flow_intro}
\end{align}

This problem appears in a number of applications that we explain in more detail below.

\subsection{Parametric Mincost-Flows in Traffic Networks}

A popular model for traffic in transportation networks is the non-atomic congestion game model due to Wardrop~\cite{wardrop1952}. In this model, we are given two designated vertices $s,t \in V$ and a flow demand of $\mu > 0$.
Every edge~$e \in E$ has a continuous and strictly increasing travel time function $f_e : \R_{\geq 0} \to \R_{\geq 0}$ that models the travel time needed to traverse the edge as a function on the total flow on that edge. A flow $\vec x$ is a \emph{Wardrop equilibrium} if it is a feasible flow and sends flow only along shortest paths, i.e., $\sum_{e \in P} f_e(x_e) \leq \sum_{e \in Q} f_e(x_e)$ for all $s$-$t$-paths $P$ and $Q$ where $x_e > 0$ for all $e \in P$. We set $\vec b = \mathbf{1}_t - \mathbf{1}_s$ where, for a node $v \in V$, we denote with $\mathbf{1}_v$ the indicator for node $v$. It is known that Wardrop equilibria coincide with the optimal solutions of the optimization problem
\begin{align}
\begin{array}{rrll}
\min & \displaystyle\sum_{e \in E} \int_0^{x_e} \mcost (\xi)\,\text{d}\xi \quad\;\;\\
\text{s.t.} & \displaystyle\sum_{e \in \delta^- (v)} x_e - \displaystyle\sum_{e \in \delta^+ (v)} x_e &= \mu b_v
	&\quad\text{for all $v \in V$}, \\
& x_e &\geq 0 & \quad\text{for all $e \in E$},
\end{array}\label{eq:wardrop_flow_intro}
\end{align}
see Beckmann et al.~\cite{beckamnn1956}. Thus, the computation of parametric minimum-cost flows \eqref{eq:mincost_flow_intro} allows us to compute Wardrop equilibria for varying flow demands. When studying transportation networks, another point of interest are system-optimal flows that minimize the overall travel time. Replacing in \eqref{eq:wardrop_flow_intro} the objective with $\sum_{e \in E} x_e \mcost(x_e)$, we obtain a flow that minimizes the total travel time of all traffic participants.

As standard measure for the inefficiency of traffic networks is the \emph{price of anarchy} first studied by Roughgarden and Tardos~\cite{roughgarden2002}. It is defined as the ratio of the total travel time of a Wardrop equilibrium and the total travel time of a system-optimal flow. There is a profound interest in understanding how the price of anarchy of traffic networks changes as the flow demand varies both (Colini-Baldeschi et al.~\cite{colinibaldeschi2020,colinibaldeschi2019}, Cominetti et al.~\cite{cominetti2019}, Englert et al.~\cite{englert2010}, O'Hare et al.~\cite{ohare2016}, Takalloo and Kwon~\cite{takalloo2020}, Wu and M\"ohring~\cite{wu2020}, Youn et al.~\cite{Youn2008}), and to improve the traffic for varying traffic demand (Christodoulou et al.~\cite{christodoulou2014}, Colini-Baldeschi et al.~\cite{colinibaldeschi2018}).

The actual computation of the price of anarchy as a function of the flow demand makes it necessary to solve two parametric mincost-flow problems as in \eqref{eq:mincost_flow_intro}, one for the Wardrop equilibrium with edge costs $\ecost(x_e) = \int_0^{x_e} \mcost(\xi) \,\text{d}\xi$ and one for the system-optimal flow with edge costs $\ecost(x_e) = x_e \mcost(x_e)$.

\subsection{Parametric Mincost-Flows in Supply Networks}

An important model for supply networks is the potential-based flow model dating back to Birkhoff and Diaz~\cite{birkhoff1956}. In this model, we are given an undirected graph $G = (V,E)$. Every edge $e \in E$ has a continuous and strictly increasing potential loss function $\mcost : \R \to \R$; usually it is additionally assumed that $\mcost(x_e) \leq 0$ for $x_e \leq 0$ and $\mcost(x_e) \geq 0$ for $x_e \geq 0$. Furthermore, we are given a balanced demand vector $\vec b$. A flow $\vec x$ is a potential-based flow, if there is a potential vector $\vec \phi = (\phi_v)_{v \in V}$ such that
\begin{align}
\label{eq:potential_equation}
\phi_u - \phi_v = \mcost(x_e) \quad \text{ for all $e = (u,v) \in E$.}
\end{align}
It is known that a flow $\vec x$ is a potential-based flow if and only if it an optimal solution of the optimization problem
\begin{align}
\begin{array}{rrll}
\min & \displaystyle\sum_{e \in E} \int_0^{x_e} \mcost (\xi)\,\text{d}\xi \quad\;\;\\
\text{s.t.} & \displaystyle\sum_{e \in \delta^- (v)} x_e - \displaystyle\sum_{e \in \delta^+ (v)} x_e &= \lambda b_v
	&\quad\text{for all $v \in V$};
\end{array}\label{eq:potential_flow}
\end{align}
see Collins et al.~\cite{collins1978} and Maugis~\cite{maugis1977}. Note that compared to \eqref{eq:wardrop_flow_intro}, we have relaxed the non-negativity constraint of the flow and, hence, the functions $\mcost$ are defined for arbitrary real values.
For potential loss functions of the form $\mcost(x_e) = \beta_e |x_e|x_e$ with $\beta_e > 0$, potential-based flows model gas flows in a pipe network; see, e.g., Weymouth~\cite{weymouth1912problems}. In this setting, the node potentials~$\phi_v$ correspond to the squared pressure at the corresponding junction in the pipe network. The parameter $\beta_e$ models different physical properties of the correspond pipe such as its diameter, its length, its slope, and the roughness of the inner wall. In a similar vein, water networks can be modeled with potential loss functions of type $\mcost(x_e) = \beta_e \text{sgn}(x_e) |x_e|^{1.852}$; see, e.g., Larock~\cite{larock2000}. Direct current (DC) power networks can be modeled with linear potential loss functions of type $\mcost(x_e) = \beta_e x_e$. Here, the parameters $\beta_e$ are equal to the conductivity of the corresponding line in the power network, and the potentials $\phi_v$ correspond to the voltage at the corresponding node, so that the potential equation~\eqref{eq:potential_equation} is equal to Ohm's law.

For the safe operation of supply networks, it is important that they are robust to varying demands. Usually, a network is safe if it obeys certain given upper and lower bounds on the potentials on the nodes and the flows along the edges. Solving parametric mincost-flows allows to compute all potential flows for a whole range of possible balance vectors, allowing to check for the safety of the resulting flows. This is particularly important for gas networks where network operators sell transmission rights that allow to send \emph{up to} a certain amount of flow between a subset of nodes. Suppose a network operator faces a base demand vector of $\vec b$ and is contemplating issuing a transmission right allowing to inject up to $\mu$ units of flow at node $s \in V$ and extract the same amount of flow from $t \in V$. Issuing this right may result in demand vectors $\vec b + \lambda \mu (\mathbf{1}_t - \mathbf{1}_s)$ with $\lambda \in [0,1]$. To determine whether all these demands can be satisfied while keeping the network safe requires the solution of a parametric minimum-cost flow problem.

\subsection{Our Results}

We study a parametric variant the minimum-cost flow problem where the inflow at the nodes is a (piecewise) linear function depending on a one-dim\-en\-sio\-nal parameter~$\lambda$. There are two main variants of this problem: A directed variant as defined in~\eqref{eq:mincost_flow_intro} and an undirected variant as defined in~\eqref{eq:potential_flow}. In prior work~\cite{KlimmWarode2021} we introduced an output-polynomial algorithm that solves the problem for the special case of piecewise linear cost functions. In this work, we consider a setting with more general, strictly convex cost functions and devise algorithms that compute approximate functions, solving the parametric minimum-cost flow problem. 

In \S~\ref{sec:functional_dependence}, we analyze the solution functions mapping the parameter~$\lambda$ to the respective minimum-cost flows and optimal potentials and compute their derivatives. Based on this theoretical foundation, in~\S~\ref{sec:parametric_computation}, we first summarize our prior work on parametric minimum-cost flows for piecewise quadratic costs and, then, develop two algorithmic approaches for the approximate, parametric computation of minimum-cost flows. 

The first approach that we call \emph{marginal cost approximation} utilizes the algorithm for piecewise quadratic cost from~\cite{KlimmWarode2021} by transforming an instance with general convex cost into an instance with piecewise quadratic cost functions. This is achieved by using a linear interpolation of the marginal cost (i.e, the derivatives of the cost functions), hence the name marginal cost approximation. We give explicit bounds on step sizes for this interpolation that guarantee a certain relative and absolute error margin for the solutions computed.
Our second approach named \emph{minimum-cost flow approximation} is based on the following, natural way for the parametric computation of minimum cost flows: Given fixed parameter values $\lambda_1, \dotsc, \lambda_K$, we compute (approximate) minimum-cost flows $\hat{\vec{x}} (\lambda_1), \dotsc, \hat{\vec{x}} (\lambda_K)$ with a given algorithm for the non-parametric minimum-cost flow problem, e.g., with the Frank-Wolfe method. In a second step, we compute a linear interpolation of these minimum cost flows and use the interpolation as an approximate solution for the original parametric problem. Again, we are able to formulate explicit bounds on the step sizes $\delta_i := \lambda_{i+1} - \lambda_{i}$ that guarantee a certain relative and absolute error margin.

Finally, in~\S~\ref{sec:computational_study}, we test both algorithmic approaches on real-world instances and evaluate their performance. To this end, we implemented MCA and MCFI in Python; the software is available as the Python package \emph{paminco} on GitHub~\cite{paminco-github}. We then generate parametric instances of various sizes based on real-world traffic and gas networks from the Transportation Network library~\cite{transportation} and the GasLib~\cite{SABHJKKIOSSS17}, respectively. The runtimes of both algorithmic approaches obtained in empirical experiments show that both algorithms, in particular the MCA method, are applicable in practice.

\subsection{Related Work}

Potential based flows have a long history as a model for gas, water, and electric networks. Birkhoff and Diaz~\cite{birkhoff1956} showed that (under reasonable assumptions) every demand vector imposes a unique flow. Similar results are obtained by Kirchhoff~\cite{kirchhoff1847} and Duffin~\cite{duffin1947}.
If the marginal cost functions $\mcost$ are linear, the non-parametric version of the mincost flow problem \eqref{eq:mincost_flow_intro} can be solved with quadratic programming techniques. In particular, there also exist an active set method due to Best~\cite{best1996} that is able to solve parametric variants of general quadratic programs. 
A standard solution method for the non-parametric variant is the Frank-Wolfe decomposition~\cite{frank1956} that is based on solving linearized version of the objective and converges to the optimal solution. Treating the problem as a linear complementarity problem it can be solved with Lemke's algorithm (e.g. Sacher~\cite{sacher1980}). The Ellipsoid method (Kozlov et al.~\cite{kozlov1980polynomial}) solves convex quadratic optimization problems in polynomial time.

The standard mincost flow problem on directed graphs with linear costs and edge capacities $u_e \in \R_{\geq 0}$ is a special case of our model where the marginal cost functions are of the form $\mcost(x) = -\infty$ for $x < 0$, $\mcost(x) = +\infty$ for $x > u_e$, and $\mcost(x) = a_e$ for some $a_e \in \R_{\geq 0}$. The standard successive shortest path algorithm implicitly solves the parametric variant of the problem.
Zadeh~\cite{zadeh1973bad} shows that the output of this parametric computation may be exponential in the input. Disser and Skutella~\cite{disser2015simplex} show that the successive shortest path algorithm is $\mathsf{NP}$-mighty, i.e., it can be used to solve $\mathsf{NP}$-hard problems while being executed.
Ahuja et al.~\cite{ahuja1984} propose an algorithm for the parametric variant of \eqref{eq:mincost_flow_intro} where the marginal cost functions are piecewise constant and show that it is output-polynomial.

\section{Preliminaries} \label{sec:preliminaries}

Let $G = (V, E)$ be a directed and strongly connected graph with vertex set $V := \{v_1,\dots,v_n\}$ and edge set $E := \{e_1,\dots, e_m\}$.
We encode $G$ by its incidence matrix $\vec \Gamma = (\gamma_{v,e}) \in \R^{n \times m}$ defined as
\begin{align*}
\gamma_{v,e} =
\begin{cases}
\phantom{-}1 &\text{ if edge $e$ enters vertex~$v$},\\
-1&\text{  if edge~$e$ leaves vertex~$v$},\\
\phantom{-}0 & \text{ otherwise}.
\end{cases}
\end{align*}
We assume that the rows of $\vec \Gamma$ are indexed $v_1,\dots,v_n$ and the columns of $\vec \Gamma$ are indexed $e_1,\dots,e_m$. We do not allow parallel edges or self loops.
For every edge $e \in E$, let $\ecost : \R_{\geq 0} \to \R_{\geq 0}$ be  a cost function. We impose the following assumptions.

\begin{assumption} For every edge~$e$, the cost function $F_e : \R_{\geq 0} \to \R_{\geq 0}$ has the following properties:
\begin{enumerate}
	\item $F_e$ is non-decreasing with $\lim_{x \to \infty} F_e(x) = \infty$;
	\item $F_e$ is strictly convex;
	\item $F_e$ is differentiable with $f_e(x) := F'_e(x)$ for all $x \geq 0$; we call $f_e$ the \emph{marginal cost function};
	\item $f_e : \R_{\geq 0} \to \R_{\geq 0}$ is piecewise differentiable, i.e., for every interval $[\alpha,\omega] \subset \R_{\geq 0}$, there are $\alpha = x_0,x_1,\dots,x_k = \omega$ with $k \in \mathbb{N}$ such that $f$ is differentiable on $(x_{i-1},x_i)$ for all $i \in \{1,\dots,k\}$.
\end{enumerate}
\end{assumption}

Let $\vec b^0 = (b_v^0)_{v \in V}$ and $\vec b = (b_v)_{v \in V} \in \R^n$ be two vectors with $\sum_{v \in V} b_v = 0$.
This paper is concerned with solving the parametric flow problem of computing a function $\vec x : [0,1] \to \R_{\geq 0}^m$ such that for all $\lambda \in [0,1]$ the flow $\vec x(\lambda)$ is an optimal solution of the optimization problem
\begin{align}
\min \sum_{e \in E} \ecost (x_e) \qquad \text{s.t.} \qquad \vec\Gamma \vec x = \vec b^0 + \lambda \vec b, \qquad \vec x \geq \vec{0},
\label{eq:mincost_flow}
\end{align}
For fixed $\lambda \in [0,1]$, the minimization problem~\eqref{eq:mincost_flow} is strictly convex and, thus, its optimal solution $\mcflowV(\lambda) = (\mcflow(\lambda))_{e \in E}$ is unique. We refer to this optimal solution as the \emph{minimum-cost flow for parameter $\lambda$}. The Karush-Kuhn-Tucker conditions imply the following optimality condition; for a proof, see, e.g., the textbook of Ruszcynski~\cite[Section~3.4]{ruszcynski2006}.
\begin{lemma} \label{lem:kkt}
For fixed $\lambda \geq 0$, let $\mcflowV = (\mcflow)_{e \in E} \in \R^m$ be a vector satisfying the constraints of~\eqref{eq:mincost_flow}. Then $\mcflowV$ is the mincost flow for parameter~$\lambda$ if and only if there exists a potential vector $\vec{\phi} = (\phi_v)_{v \in V} \in \R^n$ such that
\begin{align}
\begin{split}
\mcost (\mcflow) &= \phi_w - \phi_v
	\qquad\qquad\text{for all edges } e = (v,w) \in E \text{ with } \mcflow > 0,\\
	\mcost(\mcflow) &\geq \phi_w - \phi_v
	\qquad\qquad\text{for all edges } e = (v,w) \in E \text{ with } \mcflow = 0
.
\end{split}
\label{eq:lem:kkt}
\end{align}
\end{lemma}

We call a potential vector $\vec \phi(\lambda) \in \R^n$ \emph{optimal} if it satisfies \eqref{eq:lem:kkt} for the minimum-cost flow $\vec x^*(\lambda)$.
We are interested in the parametric solution of the optimization problem~\eqref{eq:mincost_flow}, i.e., we aim to compute a function
\begin{equation} \label{eq:mincost_flow_function}
\mcflowV : [0,1] \mapsto \R^m,
\lambda \mapsto \mcflowV (\lambda)\
,
\end{equation}
where $\mcflowV(\lambda)$ is the minimum-cost flow for parameter $\lambda$ for all $\lambda \geq 0$. Since $\mcflowV(\lambda)$ is unique, the function $\mcflowV$ is well-defined. We refer to $\mcflowV$ as the \emph{minimum-cost flow function}.
\section{Undirected minimum-cost Flows}

In this section we briefly discuss the a variant of the minimum-cost flow problem. Consider the following minimum-cost flow problem
\begin{align}
\min \sum_{e \in E} \ecost (x_e) \qquad \text{s.t.} \qquad \vec\Gamma \vec x = \vec b^0 + \lambda \vec b
\label{eq:undirected_mincost_flow}
\end{align}
without the non-negativity constraint. Since we can interpret negative flows as flow that traverses the edges in opposite direction, the flow is no longer restricted to the orientation of the edges. We therefore refer to this variant as the \emph{undirected minimum-cost flow problem}. Undirected minimum-cost flows appear in many physical real-world applications, such as electrical flows or gas flows.
Due to the missing inequality constraint, the optimality conditions from~\eqref{eq:lem:kkt} simplify to
\begin{equation} \label{eq:kkt:undirected}
\mcost (\mcflow) = \phi_w - \phi_v
	\qquad\qquad\text{for all edges } e = (v,w) \in E
.
\end{equation}
If not stated otherwise, all results in the following section hold for directed as well as undirected minimum-cost flows.

\section{Functional Dependence on $\lambda$} \label{sec:functional_dependence}

In this section, we examine the properties of the minimum-cost flow function $\vec x^* : [0, \infty) \to \R^m$ that maps the parameter $\lambda$ to the corresponding minimum-cost flow $\vec x^*(\lambda)$.
We first note that it follows from  general results on parametric convex optimization problems, e.g., Shvartsman~\cite[Theorem~2.5(i)]{shvartsman2012stability}, that the minimum-cost flow function is continuous in $\lambda$.

\begin{lemma}\label{lem:continuous}
The minimum-cost flow function $\mcflowV$ is continuous.
\end{lemma}

In order to analyze and compute the function $\mcflowV$, we proceed to use the dual information captured by the node potentials $\phi$ from the KKT-conditions in Lemma~\ref{lem:kkt}.
One difficulty, however, is that these potentials are not unique. The non-uniqueness basically stems from two reasons. First, the optimality conditions~\eqref{eq:lem:kkt} depend only on potential differences. Hence optimal potentials are invariant under additive shifts. Second, the potentials of vertices that are not connected via an edge with $x_e > 0$ may also have non-unique potentials since \eqref{eq:lem:kkt} then yields only  inequalities but no equation.

To circumvent these issues, we only consider potentials that are fixed at some arbitrarily fixed to zero vertex $v_1 \in V$. We denote by $\Pi := \{ \phi \in \R^n \mid \phi_{v_1} = 0 \}$ the set of all of these potentials and refer to $\Pi$ as the \emph{potential space}. We proceed to define a special potential function $\lambda \mapsto \vec{\pi} (\lambda)$ such that $\vec{\pi} (\lambda)$ is always an optimal potential for the minimum-cost flow for parameter $\lambda$. To this end, we consider the function
\begin{align*}
U (\lambda) &:= \big\{ e \in E : \mcflow(\lambda) > 0 \big\}
\intertext{%
that maps every $\lambda$ to the set of edges used by the minimum-cost flow for parameter $\lambda$. Since $\mcflowV$ is continuous, $U(\lambda)$ is piecewise constant with a countable set of breakpoints. For every $\lambda$, let
}
U^{-} (\lambda) &:= \{ (w,v) : (v,w) \in U(\lambda) \}
\end{align*}
be the corresponding set of backward edges. Let $\hat{E}(\lambda) := E \cup U^-(\lambda)$ and for all $e \in \hat{E}(\lambda)$,
define edge costs
\begin{align*}
\nu_e (\lambda) &=
\begin{cases}
\phantom{-} \mcost( \mcflow (\lambda)) & \text{if $e \in E$},\\
- \mcost( \mcflow (\lambda)) & \text{if $e \in U^{-} (\lambda)$}.
\end{cases}
\end{align*}
Then, the function $\vec{\pi} : \R_{\geq 0} \to \R^n$, $\lambda \mapsto \vec{\pi} (\lambda) = (\pi_v (\lambda))_{v \in V}$ as
\begin{equation}
\label{eq:potential}
\begin{split}
\pi_v(\lambda) := \, &\text{length of a shortest directed path from $v_1$ to $v$} \\ &\text{in $\hat{G}(\lambda) = (V, \hat{E}(\lambda))$ with edge cost $\nu_e(\lambda)$},
\end{split}
\end{equation}
where $v_1$ is the vertex where we fix the potential to zero. Note, that $\vec{\pi}(\lambda) \in \Pi$ for every $\lambda \geq 0$ by definition.

\begin{lemma}
\label{lem:pi-potential}
The function $\vec{\pi}$ defined in \eqref{eq:potential} is
\begin{enumerate}[(i)]
\item
	well-defined, \label{it:potential_well_defined}
\item
	piecewise continuous, \label{it:potential_piecewise_continuous}
\item
	and $\vec \pi(\lambda)$ is an optimal potential for all $\lambda \in \R_{\geq 0}$. \label{it:potential_optimal}
\end{enumerate}
\end{lemma}
\begin{proof}
We first show \textit{(\ref{it:potential_well_defined})}.
Fix $\lambda \geq 0$ arbitrarily and note that $\hat{G}(\lambda)$ is strongly connected since $G$ is strongly connected.
By Lemma~\ref{lem:kkt}, there is a potential vector $\vec\phi$ such that $\nu_e(\lambda) = \phi_w - \phi_v$ for all edges $e = (v,w) \in U(\lambda)$ and $\nu_e \geq \phi_w - \phi_v$ for all edges $e = (v,w)$ with $x_e^* = 0$. Note that this also implies that $\nu_e(\lambda) = \phi_w - \phi_v$ for all $e = (v,w) \in U^-(\lambda)$. It is straightforward to check that this implies that $\hat{G}(\lambda)$ does not contain a negative cycle. Indeed for every cycle $(v_0,v_1,\dots,v_k,v_{k+1})$ with $v_{k+1} = v_0$ we obtain
\begin{align*}
\sum_{i=0}^k \nu_{(v_i, v_{i+1})}(\lambda) \geq \sum_{i=0}^{k} \phi_{v_{i+1}} - \phi_{v_i}	= \phi_{v_{k+1}} - \phi_{v_0} = 0.
\end{align*}
Since $\hat{G}(\lambda)$ is strongly connected and does not contain a negative cycle, the shortest path lengths exist and the values $\phi_v(\lambda)$ are well-defined.

We proceed to show \textit{(\ref{it:potential_piecewise_continuous})}.
The sets $U(\lambda)$ are piecewise constant, implying the edge set of $\hat{G}(\lambda)$ is piecewise constant. Since the marginal costs $\mcost$ and the  optimal flows $\mcflow$ are continuous, so are the edge cost $\nu_e$. Hence, the shortest path length are continuous as long as the graph is unchanged and, therefore, the function $\vec{\pi}$ is piecewise continuous.

Finally, we show \textit{(\ref{it:potential_optimal})}.
To show that $\vec \pi$ is an optimal potential, we show that the conditions from~\eqref{eq:lem:kkt} are satisfied. Since $\vec \pi$ is a shortest path potential, the inequality constraints in~\eqref{eq:lem:kkt} are satisfied for all edges. Since these inequalities must hold in both directions for edges in $U(\lambda)$, the equality constraints follow as well.
\end{proof}

The minimum-cost flow $\mcflow(\lambda)$ depends continuously on the optimal potentials via the equalities in~\eqref{eq:lem:kkt}. In particular, the flow can be expressed as a solution to the system of equations in~\eqref{eq:lem:kkt}. Therefore, we are interested in the edges where the inequality in the optimality conditions are satisfied with equality. We denote this edge set by
\[
S(\lambda) := \big\{ e = (v,w) \in E \;:\; \mcost(\mcflow (\lambda)) = \pi_w (\lambda) - \pi_v (\lambda) \big\}
\]
and call these edges \emph{support}. Note that $U(\lambda) \subseteq S(\lambda)$ but $S(\lambda)$ may contain more edges than $U(\lambda)$. The support $S(\lambda)$ is still piecewise constant since the edge cost $\nu_e(\lambda)$ are continuous and $U(\lambda)$ is piecewise constant. We proceed to argue that the graph with edge set $S(\lambda)$ is also connected for all $\lambda \geq 0$.

\begin{corollary}
\label{cor:support-connected}
For every $\lambda > 0$, the graph $G = (V, S(\lambda))$ is connected.
\end{corollary}

\begin{proof}
By construction, $\pi_v(\lambda)$ is equal to the length of a shortest directed path from $v_1$ to $v$ in $\hat{G}(\lambda)$. As argued in the proof of Lemma~\ref{lem:pi-potential}, the graph $\hat{G}(\lambda)$ is strong connected, thus, there exists a shortest path from $v_1$ to $v$ for all $v \in V$. All edges on this shortest path are contained in $S(\lambda)$, and the result follows.
\end{proof}

In the following, it will be convenient to consider vertices and matrices without columns and rows corresponding to the vertex $v_1$. For a vector $\vec{z} \in \R^n = (z_1,z_2,\dots,z_n)^\top$, denote by $\hat{\vec{z}} = (z_2,\dots,z_n)^\top$ the vector with the first entry (corresponding to $v_1$) removed.
Similarly, for a matrix $\vec{A} \in \R^{n \times n}$, denote by $\hat{\vec{A}}$ the matrix with the first row and column removed.
If $\hat{\vec{A}}^{-1}$ exists, we define the matrix
\[
\vec{A}^* :=
\begin{bmatrix}
0 & \vec{0}^{\top} \\
\vec{0} & \hat{\vec{A}}^{-1}
\end{bmatrix}
.
\]
This matrix satisfies $\vec{A}\vec{A}^* \vec{A} = \vec{A}$ and, hence, is the unique generalized inverse mapping into the subspace $\{ \vec{z} \in \R^n \;|\; z_{v_1} = 0 \}$ of vectors where the component corresponding to $v_1$ is zero.

With these definitions, we are now in position to obtain explicit formulas for the derivatives of the functions $\lambda \mapsto \vec x(\lambda)$ and $\lambda \mapsto \vec \pi(\lambda)$ that map the parameter~$\lambda$ to the minimum-cost flow and optimal potential, respectively.

\begin{theorem} \label{thm:derivative}
Assume that the marginal cost functions $\mcost$ are differentiable with $\mcost(x) > 0$ for all $x \geq 0$ and $e \in E$.
Let $I \subset [0, \infty)$ be an open interval such that the support is constant on $I$, i.e., $S (\lambda) = S$ for all $\lambda \in I$. Then, the functions $\lambda \mapsto \vec{x}^* (\lambda)$ and $\lambda \mapsto \vec{\pi} (\lambda)$ are differentiable on $I$ with the explicit derivatives
\begin{align}
\frac{d}{d \lambda} \vec{\pi} (\lambda) &= \vec{L}_{\lambda}^* \vec{b} & &\text{and} &
\frac{d}{d \lambda} \vec{x}^* (\lambda) &= \vec{C}_{\lambda} \vec{\Gamma}^{\top} \vec{L}_{\lambda}^*
 \vec{b}
 ,
\label{eq:thm:derivative}
\end{align}
where
$\vec{C}_{\lambda} = \diag \big(c_{e_1} (\lambda), \dotsc, c_{e_m} (\lambda) \big)$
with
\begin{align*}
c_e (\lambda) :=
\begin{cases}
0
	&\text{if } e \notin S(\lambda), \\
\frac{1}{\mcost' (x^*_e (\lambda))}
	&\text{if } e \in S(\lambda)
\end{cases}
\end{align*}
and $\vec{L}_{\lambda} = \vec{\Gamma} \vec{C}_{\lambda} \vec{\Gamma}^{\top}$.
\end{theorem}

\begin{proof}
Let $S$ be the (constant) support on the interval $I$. We encode this support via the matrix $\vec{S} := \diag (s_{e_1}, \dotsc, s_{e_m})$, where $s_e = 1$ if $e \in S$ and $s_e = 0$ otherwise. Further, we denote by $\vec{\Gamma}_{-1}$ the submatrix of the incidence matrix obtained by removing the first row.

The proof is based on the following idea. We define a function $\vec{g}$ that maps a triplet $(\vec{x}, \hat{\vec{\phi}}, \lambda)$ consisting of a flow vector, a (shortened) potential vector and a parameter value to the slack of the equalities in the KKT conditions from~\eqref{eq:lem:kkt}. We then show that whenever $\vec{g} \big( \vec{x}, \hat{\vec{\phi}}, \lambda \big) = \vec{0}$, $\vec{x}$ is a minimum-cost flow for parameter~$\lambda$ with optimal potential $\vec{\phi}$. The implicit function theorem then yields differentiable functions mapping the parameter to minimum-cost flows and optimal potentials. By uniqueness, these functions must coincide with the minimum-cost flow function and the optimal potential function, concluding the proof.

Formally, we define a function $\vec{h} \colon \R^m \to \R^m, \vec{x} \mapsto \vec{h} (\vec{x}) = h_e \big( x_e(\lambda)\big)_{e \in E}$ with $h_e (x_e) = \mcost(x_e)$ if $e \in S(\lambda)$ and $h_e (x_e) = x_e$ if $e \notin S(\lambda)$. Denote by $\vec{A}$ the Jacobian matrix of the function $\vec{h}$. It is easy to see that $\vec{A}$ is a diagonal matrix with coefficients $a_e = \mcost' (x_e)$ if $e \in S(\lambda)$ and $a_e = 1$ otherwise.

Then, we define the function
\[
\vec{g} \colon \R^m \times \R^{n-1} \times I \to \R^{m+n-1}
\quad\text{with}\quad
(\vec{x}, \hat{\vec{\phi}}, \lambda) \mapsto
\begin{bmatrix}
\vec{h}(\vec{x}) - \vec{S} \vec{\Gamma}_{-1}^{\top} \hat{\vec{\phi}} \\
\vec{\Gamma}_{-1} \vec{x} - \vec{b}^0 - \lambda \vec{b}
\end{bmatrix}
\]
and claim that for every $\lambda \in I$, $\vec{x} \in R^m$, and $\vec{\phi} \in \R^n$ with $\pi_1 = 0$,
\begin{equation} \label{eq:prf:derivative:1}
\vec{g} (\vec{x}, \hat{\vec{\phi}}, \lambda) = \vec{0}
\end{equation}
implies that $\vec{x} = \vec{x}^*(\lambda)$, i.e., $\vec{x}$ is the minimum-cost flow for parameter~$\lambda$, and $\vec{\phi} = \vec{\pi}(\lambda)$, i.e., $\vec{\phi}$ is the optimal potential for this flow.
Indeed, we observe that the last $n-1$ lines of \eqref{eq:prf:derivative:1} are equivalent to the flow conservation on all vertices but $v_1$. Since the demands sum to $1$ this also implies the flow conservation for the whole graph.
Now consider any of the first $m$ rows of~\eqref{eq:prf:derivative:1}. Each of these rows corresponds to an edge~$e = (v, w) \in E$ and  is equivalent to $h_e (x_e) - s_e (\pi_w - \pi_v) = 0$. Thus, the first $m$ rows of~\eqref{eq:prf:derivative:1} can be restated as
\begin{align*}
\mcost(x_e) &= \pi_w - \pi_v \phantom{0}
	\qquad \text{if } e = (v, w) \in S, \\
x_e &= 0 \phantom{\pi_w - \pi_v}
	\qquad \text{if } e = (v, w) \notin S.
\end{align*}
This implies that the flow $\vec{x}$ restricted to the subgraph $G' = (V, S)$ satisfies the optimality conditions from Lemma~\ref{lem:kkt} on this subgraph. Hence, the resitriction of $\vec{x}$ is a minimum-cost flow on $G'$. Since the minimum-cost flow does not change if all non-support edges are removed, $\vec{x}$ must coincide with the minimum-cost flow for the whole graph on $G'$. Since, further the minimum-cost flow as well as $\vec{x}$ are zero on all non-support edges, we conclude that $\vec{x}$ coincides with the minimum-cost flow.
Further, we observe that the potential differences $\phi_w - \phi_v = \mcost(x_e) = \mcost(x^*_e (\lambda) = \pi_v (\lambda) - \pi_v(\lambda)$ coincide for all edges $e = (v,w)$ in the support~$S$. Since $\phi_{v_1} = \pi_{v_1} = 0$ and $G' = (V,S)$ is connected, we get that also the potentials $\vec{\phi}$ and $\vec{\pi} (\lambda)$ coincide.

We now compute the Jacobian matrix of the function $\vec{g}$
\begin{align*}
\vec{J}_{\vec{g}} (\vec{x}, \hat{\vec{\phi}}, \lambda) &=
\begin{bmatrix}
\vec{A} & \quad- \vec{S} \vec{\Gamma}_{-1}^{\top} & \phantom{-}\vec{0} \\
\vec{\Gamma}_{-1} & \vec{0} &\quad -\hat{\vec{b}}
\end{bmatrix}
\end{align*}
consisting of the submatrices
\[
\vec{J}_{\vec{g}, (\vec{x}, \hat{\vec{\phi}})} =
\begin{bmatrix}
\vec{A} & \quad - \vec{S} \vec{\Gamma}_{-1}^{\top} \\
\vec{\Gamma}_{-1} & \vec{0}
\end{bmatrix}
\quad \text{and} \quad
\vec{J}_{\vec{g}, \lambda} =
\begin{bmatrix}
\phantom{-}\vec{0} \\
-\hat{\vec{b}}
\end{bmatrix}
\]
The matrix $\vec{J}_{\vec{g}, (\vec{x}, \hat{\vec{\phi}})}$
is non-singular if and only if $\hat{\vec{L}}_{\lambda} := \vec{\Gamma}_{-1} \vec{A}^{-1} \vec{S} \vec{\Gamma}_{-1}^{\top} = \vec{\Gamma}_{-1} \vec{C}_{\lambda} \vec{\Gamma}_{-1}^{\top}$ is non-singular; see, e.g., the textbook by Harville~\cite[Theorem~8.5.11]{harville1997matrix}. The matrix $\hat{\vec{L}}_{\lambda}$ is obtained from the matrix $\vec{L}_{\lambda} = \vec{\Gamma} \vec{C}_{\lambda} \vec{\Gamma}^{\top}$ by deleting the first row and column. The latter matrix $\vec{L}_{\lambda}$ is the weighted Laplacian matrix of the subgraph induced by all edges $e \in S(\lambda_0)$ in the support with respect to the weights $c_{e} (x_e)$.
The subgraph induced by the edges in the support is connected and basic theory of Laplacian matrices (see, e.g., \cite{gre1991geometry,merris1994laplacian,mohar1991laplacian}) implies that $\vec{L}_{\lambda}$ has rank $n-1$ and $\hat{\vec{L}}_{\lambda} = \vec{\Gamma}_{-1}\vec{C}_{\lambda} \vec{\Gamma}_{-1}^{\top}$ is non-singular. Thus, again by Harville~\cite[Theorem~8.5.11]{harville1997matrix}, the matrix $\vec{J}_{g, (\vec{x}, \hat{\vec{\phi}})}$ is non-singular with inverse
\begin{align*}
\vec{J}^{-1}_{g, (\vec{x}, \hat{\vec{\phi}})} =
\begin{bmatrix}
\vec{A}^{-1} - \vec{A}^{-1} \vec{S} \vec{\Gamma}_{-1}^{\top} \hat{\vec{L}}_{\lambda}^{-1} \vec{\Gamma}_{-1} \vec{A}^{-1} & \quad\vec{A}^{-1} \vec{S} \vec{\Gamma}_{-1}^{\top} \hat{\vec{L}}_{\lambda}^{-1} \\
- \hat{\vec{L}}_{\lambda}^{-1} \vec{\Gamma}_{-1} \vec{A}^{-1} & \hat{\vec{L}}_{\lambda}^{-1}
\end{bmatrix}
.
\end{align*}
By the implicit function theorem, there is an open set $U \ni \lambda_0$ such that there exist continuous and differentiable functions $\vec{x} : U \to \R^m$ and $\hat{\vec{\phi}} : U \to \R^n$ such that $g( \vec{x} (\lambda), \hat{\vec{\phi}} (\lambda), \lambda) = \vec{0}$ for all $\lambda \in U$. As argued above, $\vec{x}(\lambda) = \mcflowV (\lambda)$ and $\vec{\phi} (\lambda) = \vec{\pi} (\lambda)$.  Thus, the minimum-cost flow function $\mcflowV (\lambda)$ and the potential function $\vec{\pi} (\lambda)$ are differentiable for almost all $\lambda \geq 0$.

Finally, the implicit function theorem and the chain rule imply that
\begin{align*}
\frac{d}{d \lambda} \begin{bmatrix} \vec{x} (\lambda) \\ \hat{\vec{\pi}} (\lambda) \end{bmatrix}
= - \vec{J}^{-1}_{g, (\vec{x}, \hat{\vec{\phi}})} \, \vec{J}_{g, \lambda} =
\begin{bmatrix}
\vec{A}^{-1} \vec{S} \vec{\Gamma}_{-1}^{\top} \hat{\vec{L}}_{\lambda}^{-1} \hat{\vec{b}} \\
\hat{\vec{L}}_{\lambda}^{-1} \hat{\vec{b}}
\end{bmatrix}.
\end{align*}
This implies
\begin{align*}
\frac{d}{d \lambda} \vec x(\lambda) &= \vec A^{-1} \vec S \vec{\Gamma}_{-1}^\top \hat{\vec L}_{\lambda}^{-1} \hat{\vec b} = \vec C_{\lambda}	\begin{bmatrix} \vec 0 & \vec{\Gamma}_{-1}^\top \end{bmatrix} \begin{bmatrix} 0 & \vec  0^\top \\ \vec 0 & \hat{\vec L}_{\lambda}^{-1}\end{bmatrix} \begin{bmatrix} b_{v_1}\\ \hat{\vec b} \end{bmatrix} = \vec C_{\lambda} \vec \Gamma^{\top} \vec L_{\lambda}^* \vec b,\\
\frac{d}{d \lambda} \vec \pi(\lambda)  &= \begin{bmatrix} 0 \\ \frac{d}{d\lambda} \hat{\vec \pi}(\lambda) \end{bmatrix} = \begin{bmatrix} 0 & \vec 0^\top \\ \vec 0 & \hat{\vec L}^{-1}_{\lambda} \end{bmatrix} \begin{bmatrix} b_{v_1} \\ \hat{\vec b} \end{bmatrix},
\end{align*}
concluding the proof.
\end{proof}

We can use the result from Theorem~\ref{thm:derivative} for two purposes. First, we notice that if all cost functions are quadratic, the matrices $\vec{C}_{\lambda}$ and $\vec{L}_{\lambda}$ are independent of $\lambda$ since $\mcost'$ is constant for all edges in this case. Hence, the derivatives of the potentials $\vec{\pi}$ and the flows $\vec{x}$ are constant as long as the support does not change and the functions $\vec{\pi}$ and $\vec{x}$ are piecewise linear functions that can be computed directly using the Laplacian matrix $\vec{L}$ given a fixed support. This insight can be used to develop a homotopy method to compute the functions $\vec{\pi}$ and $\vec{x}$ by solving these Laplacian systems and performing pivoting steps to obtain the correct supports. We describe this method that also works for piecewise-quadratic cost functions in more detail in~\S~\ref{sec:quadratic}.

Second, the result from Theorem~\ref{thm:derivative} can be used to obtain closed formulas for the derivatives of the objective function with respect to the parameter $\lambda$. To this end, denote by
\[
C (\lambda) := \sum_{e \in E} \ecost (x_e (\lambda))
\]
the cost of the mincost flow $\vec{x}(\lambda)$ for parameter $\lambda$. We then obtain the following result.
\begin{theorem} \label{thm:derivative_objective}
Assume that the marginal cost functions are differentiable and $\mcost (x) > 0$ for all $x \geq 0$ and all $e \in E$. Then,
\begin{enumerate}[(i)]
\item
	the function $\lambda \mapsto C(\lambda)$ is differentiable for all $\lambda \geq 0$ with
	\begin{align*}
	\frac{d}{d \lambda} C (\lambda) &= \vec{b}^{\top} \vec{\pi} (\lambda)
	.
	\end{align*}
	If the demand function is linear, i.e., $\vec{b}^0 = \vec{0}$, and non-decreasing marginal cost functions $\mcost (x)$, we additionally have $\frac{d}{d \lambda} C (\lambda) \geq 0$.
\item
	the function $\lambda \mapsto C(\lambda)$ is twice differentiable on every interval $I$ where the support is constant with
\begin{align*}
\frac{d^2}{d^2 \lambda} C (\lambda) &= \vec{b}^{\top} \vec{L}^*_{\lambda} \vec{b} > 0
.
\end{align*}
\end{enumerate}
\end{theorem}
\begin{proof}
Let $\lambda \geq 0$. Then, there are two cases.

Case 1:
	The support is constant on an open interval containing $\lambda$. (Since the support function is piecewise constant, this is the case for almost all $\lambda \geq 0$.) In this case, we can apply Theorem~\ref{thm:derivative} and the derivatives $\frac{d}{d \lambda} \vec{\pi} (\lambda)$ and $\frac{d}{d \lambda} \vec{x} (\lambda)$ exist. We note that for all edges $e \in S(\lambda)$ in the support, $\mcost (x_e (\lambda)) = \pi_w (\lambda) - \pi_v (\lambda)$ and for all edges $e \notin S(\lambda)$ not in the support, $\frac{d}{d \lambda} x_e (\lambda) = 0$ by~\eqref{eq:thm:derivative}. We then compute
	\begin{align*}
	\frac{d}{d \lambda} C (\lambda) &= \sum_{e \in E} \mcost (x_e (\lambda)) \frac{d}{d \lambda} x_e (\lambda) \\
	&=
	\sum_{e = (v,w) \in E} \big( \pi_w (\lambda) - \pi_v (\lambda) \big) \frac{d}{d \lambda} x_e (\lambda) \\
	&= \big( \vec{\Gamma}^{\top} \vec{\pi} (\lambda) )^{\top} \frac{d}{d \lambda} \vec{x}(\lambda) \\
	&= \big( \vec{\pi} (\lambda) \big)^{\top} \frac{d}{d \lambda} \vec{\Gamma} \vec{x}(\lambda)  = \big( \vec{\pi} (\lambda) \big)^{\top} \vec{b}.
	\end{align*}
	Again by Theorem~\ref{thm:derivative}, we also obtain
	$\frac{d^2}{d^2 \lambda} C (\lambda) = \vec{b}^{\top} \vec{L}^*_{\lambda} \vec{b}$.
	Since $\vec{L}_{\lambda}$ is a Laplacian matrix with positive edge weights, its submatrix $\hat{\vec{L}}_{\lambda}$ is strictly positive definite. Therefore, with the definition of the pseudo inverse, we obtain
	$\frac{d^2}{d^2 \lambda} C (\lambda) = \vec{b}^{\top} \vec{L}^*_{\lambda} \vec{b} = \hat{\vec{b}}^{\top} \hat{\vec{L}}^*_{\lambda} \hat{\vec{b}} > 0$.

Case 2:
	The support function has a breakpoint in $\lambda$. Then the derivatives $\frac{d}{d \lambda} \vec{\pi} (\lambda)$ and $\frac{d}{d \lambda} \vec{x} (\lambda)$ may not exist anymore. However, since the support function is piecewise constant, there is an open interval $U \ni \lambda$, such that the derivatives exists for all $\lambda \neq \tilde{\lambda} \in U$. Since the potential function $\vec{\pi} (\lambda)$ is continuous, we get that $\lim_{\tilde{\lambda} \to \lambda} \frac{d}{d \lambda} C (\tilde{\lambda}) = \big( \vec{\pi} (\lambda) \big)^{\top} \vec{b}$ exists. Since the function $\lambda \mapsto C(\lambda)$ is continuous, this implies that it is also differentiable in all $\lambda$.

Finally, consider the special case of a linear demand function and non-decreasing marginal cost functions. For every $\lambda > 0$, we compute
\begin{align*}
\frac{d}{d \lambda} C(\lambda) &= \frac{1}{\lambda} \big( \vec{\Gamma} \vec{x} (\lambda) \big)^{\top} \vec{\pi} (\lambda) = \frac{1}{\lambda} \big(\vec{x} (\lambda) \big)^{\top} \vec{\Gamma}^{\top} \vec{\pi} (\lambda) \\
&= \frac{1}{\lambda} \sum_{e \in E} \big( \pi_w (\lambda) - \pi_v ( \lambda )  \big) x_e ( \lambda )
= \frac{1}{\lambda} \sum_{e \in E} \mcost (x_e (\lambda))  x_e ( \lambda ) \geq 0
.
\end{align*}
By continuity, $\frac{d}{d \lambda} C(\lambda) \geq 0$ follows for all $\lambda \geq 0$.
\end{proof}

Finally, we state a theorem with two properties of Laplacian matrices, that are in particular known in the context of electrical networks, where these properties are sometimes called Thomson's principle and Rayleigh's monotonicity law (see, e.g.,~\cite{doyle1984random}).

\begin{theorem} \label{thm:thomson_rayleigh}
Let $\vec{x} (\lambda)$ be the mincost flow solving~\eqref{eq:mincost_flow} for some $\lambda \geq 0$. For every edge $e \in E$, let $a_e := \mcost' (x_{e} (\lambda))$. Define the matrix $\vec{A}_{\lambda} := \diag \big(a_{e_1}, \dotsc, a_{e_m} \big)$.
\begin{enumerate}[(i)]
\item \label{it:thomson_principle} {\bf Thomson's principle:}
	For every feasible flow $\vec{y}$ for demand $\lambda \vec{b}$ with support $S(\lambda)$, i.e., $y_e > 0$ only if $e \in S(\lambda)$, we have
	\[
	\bigg( \frac{d}{d \lambda} \vec{x} (\lambda) \bigg)^{\top} \vec{A}_{\lambda}  \frac{d}{d \lambda} \vec{x} (\lambda) \leq \vec{y}^{\top} \vec{A}_{\lambda} \vec{y}
	.
	\]
\item \label{it:rayleighs_law} {\bf Rayleigh's monotonicity law:}
	Let
	$
	\tilde{\vec{C}} = \diag\Big(
	\frac{\mathds{1}_{e_1 \in S(\lambda)}}{\tilde{a}_1}
	, \dotsc,
	\frac{\mathds{1}_{e_m \in S(\lambda)}}{\tilde{a}_m}
\Big)$ with $\tilde{a}_e \geq a_e$ for all edges~$e$, and let $\tilde{\vec{L}} = \vec{\Gamma} \tilde{\vec{C}} \vec{\Gamma}^{\top}$. Then,
	\[
	\vec{b}^{\top} \vec{L}^*_{\lambda} \vec{b} \leq
	\vec{b}^{\top} \tilde{\vec{L}}^* \vec{b}
	.
	\]
\end{enumerate}
\end{theorem}
\begin{proof}
We observe that $\vec{A} \vec{C}_{\lambda} = \diag\big( \mathds{1}_{e_1 \in S(\lambda)}, \dotsc,  \mathds{1}_{e_m \in S(\lambda)} \big) =: \vec{J}$.
For every edge $e$, let $\vec{z} := \vec{y} - \frac{d}{d \lambda} \vec{x} (\lambda)$. Then, $\vec{\Gamma} \vec{z} = \vec{0}$ and, since $x_e = y_e = 0$ for all $e \notin S(\lambda)$, $\vec{z} \vec{J} = \vec{z}$. Overall, we get
\begin{align*}
\vec{y}^{\top} \vec{A}_{\lambda} \vec{y}
&=  \bigg( \frac{d}{d \lambda} \vec{x} (\lambda) \bigg)^{\top} \vec{A}_{\lambda}  \frac{d}{d \lambda} \vec{x} (\lambda) + 2 \, \vec{z}^{\top} \vec{A}_{\lambda} \frac{d}{d \lambda} \vec{x} (\lambda) + \vec{z} \vec{A}_{\lambda} \vec{z} \\
&\geq
\bigg( \frac{d}{d \lambda} \vec{x} (\lambda)\bigg)^{\top} \vec{A}_{\lambda} \frac{d}{d \lambda} \vec{x} (\lambda) + 2 \, \vec{z}^{\top} \vec{A}_{\lambda} \vec{C}_{\lambda} \vec{\Gamma}^{\top} \vec{L}_{\lambda} \vec{b} \\
&= \bigg( \frac{d}{d \lambda} \vec{x} (\lambda) \bigg)^{\top} \vec{A}_{\lambda}  \frac{d}{d \lambda} \vec{x} (\lambda) + 2\, \big( \vec{\Gamma} \vec{J} \vec{z} \big)^{\top} \vec{L}_{\lambda} \vec{b}
= \bigg( \frac{d}{d \lambda} \vec{x} (\lambda) \bigg)^{\top} \vec{A}_{\lambda} \frac{d}{d \lambda} \vec{x} (\lambda)
.
\end{align*}

For \emph{(\ref{it:rayleighs_law})}, let $\vec{y} = \tilde{\vec{C}} \vec{\Gamma}^{\top} \tilde{\vec{L}}^* \vec{b}$. Then $\vec{\Gamma}^{\top} \vec{y} = \vec{b}$ (i.e., $\vec{y}$ is a flow for demand $\vec{b}$). Let $\tilde{\vec{A}} := \diag (\tilde{a}_{e_1}, \dotsc, \tilde{a}_{e_m})$. Then
\begin{align*}
\vec{b}^{\top} \vec{L}^*_{\lambda} \vec{b}
&= \bigg( \frac{d}{d \lambda} \vec{x} (\lambda) \bigg)^{\top} \vec{\Gamma}^{\top} \vec{L}^*_{\lambda} \vec{b}
= \bigg( \frac{d}{d \lambda} \vec{x} (\lambda) \bigg)^{\top} \vec{A} \vec{C}_{\lambda} \vec{\Gamma} \vec{L}^*_{\lambda} \vec{b}  \\
&= \bigg( \frac{d}{d \lambda} \vec{x} (\lambda) \bigg)^{\top} \vec{A} \frac{d}{d \lambda} \vec{x} (\lambda)
\stackrel{\mathclap{\text{\emph{(\ref{it:thomson_principle})}}}}{\leq}
\vec{y}^{\top} \vec{A} \vec{y}
\leq \vec{y}^{\top} \tilde{\vec{A}} \vec{y}
= \vec{b}^{\top} \tilde{\vec{L}}^* \vec{b},
\end{align*}
where the last equality follows with the same steps reversed.
\end{proof}

\section{Parametric Computation of Minimum-Cost Flows}
\label{sec:parametric_computation}

Following the theoretical analysis in the previous section, we now want to develop methods for the computation of the minimum-cost flow function. As it can be seen from the optimality conditions~\eqref{eq:lem:kkt}, every minimum-cost flow for a fixed parameter is a solution to a system of equalities. If these equations are linear with rational coefficients, the solution is also rational and can be computed explicitly. If these equations are non-linear, the solutions may be irrational and we therefore can only compute approximate solutions.
Therefore, this sections consists of two parts.
First, we consider the case of parametric minimum-cost flows in networks with (piecewise) quadratic cost functions. In this case, the marginal costs, and thus the optimality conditions from~\eqref{eq:lem:kkt}, are (piecewise) linear. We then can use a homotopy method based on the computation of electrical flows that we introduced in a previous paper~\cite{KlimmWarode2021} in order to compute the exact minimum-cost flow functions.
In the second part, we consider more general, convex cost functions and develop two methods for the computation of approximate minimum-cost flow functions. The first method that we call \emph{minimum-cost approximation} (MCA) is based  on a linear interpolation of the non-linear marginal cost functions allowing us to apply the algorithm developed in the first part.
The second method called \emph{minimum-cost flow interpolation} (MCFI) relies on the computation of minimum-cost flows for fixed demands with a suitable method, e.g., the Frank-Wolfe algorithm. The approximate minimum-cost flow function computed by this method is a linear interpolation of fixed demand minimum-cost flows at fixed breakpoints.
For both MCA and MCFI we develop bounds for the step sizes of the respective interpolations that guarantee bounds on the error of the approximation.

\subsection{Minimum-cost Flows With Piecewise Quadratic Cost}
\label{sec:quadratic}

We assume throughout this section that all cost functions $\ecost (x)$ are piecewise quadratic. More specifically, we assume that for every edge there exists a piecewise linear \emph{marginal cost function} $\mcost : \R \to \R$ such that $\ecost (x) = \int_0^x \mcost(s) ds$. Then the derivatives of the marginal costs, $\mcost' (x)$ are piecewise constant. Since the Laplacian matrix $\vec{L}_{\lambda}$ defined in Theorem~\ref{thm:derivative} only depends on this second derivative, the matrix $\vec{L}_{\lambda}$ and therefore the derivatives of the mincost flow function and the potential function are piecewise constant as well. Thus the mincost flow function $\lambda \mapsto \vec{x} (\lambda)$ is piecewise linear. The aim of this subsection is to develop an algorithm that computes these piecewise linear functions explicitly.

\subsubsection{Undirected parametric minimum-cost flows}

We start by considering the undirected parametric minimum-cost flow problem from~\eqref{eq:undirected_mincost_flow}. In the case of linear marginal cost functions, i.e., if $\mcost( x) = a_e x$, then the optimality conditions~\eqref{eq:kkt:undirected} for this undirected problem are $a_e x_e = \pi_w - \pi_v$ for all edges $e = (v,w) \in E$. If we interpret the potential difference as voltage, the flow as electrical current, and the slope of the marginal cost function as resistance, then this optimality condition is exactly \emph{Ohm's law}. Together with the primal flow conservation constraint, we see that an undirected minimum-cost flow with respect to linear marginal costs satisfies the conditions for an electrical flow in a linear resistor network. We therefore also refer to this setting as the \emph{electrical flow setting}.

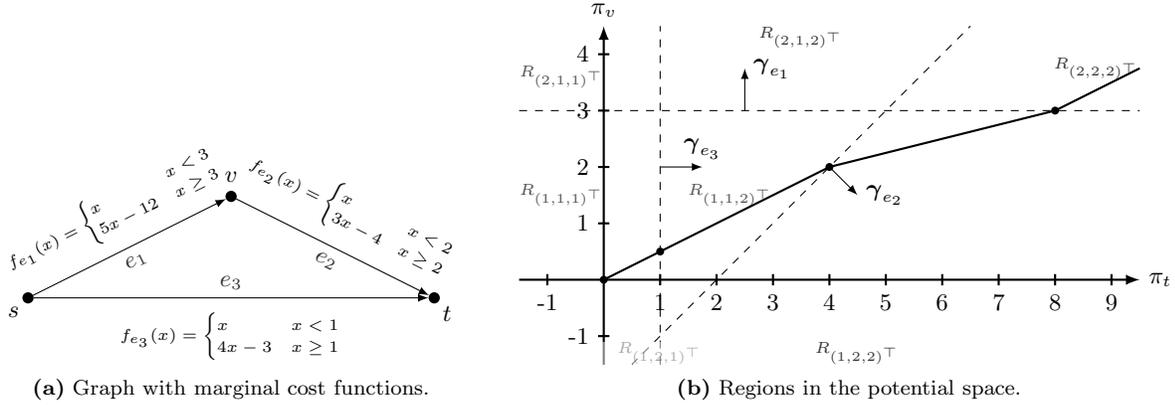
\begin{figure}[t]
\centering
\begin{minipage}[b]{.35\linewidth}
\begin{center}
\footnotesize
\begin{tikzpicture}[scale=.9]
\useasboundingbox
	(0,-1) rectangle (6,2);

\draw (0,0) node[solid] (s) {} node [below left] {$s$};
\draw (3,1.5) node[solid] (v) {} node [above=.25em] {$v$};
\draw (6,0) node[solid] (t) {} node [below right] {$t$};

\draw[->]	(s) edge node[midway, below, sloped,black!80] {$e_1$} node[midway, above=0.25em, sloped] {\tiny $\mcost[e_1] (x) = \begin{cases} x & x< 3 \\ 5x - 12 & x \geq 3\end{cases}$} (v)
			(v) edge node[midway, below, sloped,black!80] {$e_2$} node[midway, above=0.25em, sloped] {\tiny $\mcost[e_2] (x) = \begin{cases} x & x< 2 \\ 3x - 4 & x \geq 2 \end{cases}$} (t)
			(s) edge node[midway, above,black!80] {$e_3$}  node[midway, below=0.25em] {\tiny $\mcost[e_3] (x) = \begin{cases} x & x< 1 \\ 4x - 3 & x \geq 1\end{cases}$} (t);
\end{tikzpicture}
\\
{\scriptsize \textbf{(a)} Graph with marginal cost functions.}
\end{center}
\end{minipage}
\begin{minipage}[b]{.63\linewidth}
\begin{center}
\footnotesize
\begin{tikzpicture}[scale=0.75]
\newcommand{\xMin}{-1.5} \newcommand{\xMax}{9.5}
\newcommand{\yMin}{-1.5} \newcommand{\yMax}{4.5}

\draw[thick, ->] (\xMin,0) -- (\xMax,0) node[anchor=west] {$\pi_t$};
\foreach \i in {-1,1,2,...,9}
{
	\draw[thick] (\i, 0.1) -- (\i,-0.1) node[anchor=north] {\i};
}
\foreach \i in {-1,1,2,3,4}
{
	\draw[thick] (0.1, \i) -- (-0.1, \i) node[anchor=east] {\i};
}
\draw[thick, ->] (0,\yMin) -- (0,\yMax) node[anchor=south] {$\pi_v$};

\draw[dashed] (1, \yMin) -- (1, \yMax);
\draw[dashed] (\xMin,3) -- (\xMax,3);
\draw[dashed] ({max(\xMin,\yMin+2)},{max(\xMin-2,\yMin)}) -- ({min(\xMax,\yMax+2)},{min(\xMax-2,\yMax)});

\node[anchor=south, inner sep=0pt, black!80] at (4.5,\yMin) {\tiny $\region[(1,2,2)^{\top}]$};
\node[anchor=east, inner sep=0pt, black!80] at (\xMax,3.75) {\tiny $\region[(2,2,2)^{\top}]$};
\node[anchor=north, inner sep=0pt, black!80] at (3.5,\yMax) {\tiny $\region[(2,1,2)^{\top}]$};
\node[anchor=north west, black!70, inner sep=0pt]
	at (-1.5,3.8) {\tiny $\region[(2,1,1)^{\top}]$};
\node[anchor=north west, black!70, inner sep=0pt]
	at (-1.5,1.7) {\tiny $\region[(1,1,1)^{\top}]$};
\node[black!70, inner sep=0pt] at (2.25,1.5) {\tiny $\region[(1,1,2)^{\top}]$};
\node[anchor=south, black!70, fill=white, fill opacity=.5, text opacity=1, inner sep=0pt]
	at (0.85,\yMin) {\tiny $\region[(1,2,1)^{\top}]$};

\draw[thin, ->] (2.5,3) -- (2.5, 3.75) node[right] {$\vec{\gamma}_{e_1}$};
\draw[thin, ->] (1,2) -- (1.75, 2) node[above] {$\vec{\gamma}_{e_3}$};
\draw[thin, ->] (4,2) -- (4.5, 1.5) node[right] {$\vec{\gamma}_{e_2}$};


\draw[thick]
	(0,0) -- (4,2) -- (8,3) -- ++(3/2,3/4);

\fill (0,0) circle(2pt);
\fill (1,1/2) circle(2pt);
\fill (4,2) circle(2pt);
\fill (8,3) circle(2pt);


\end{tikzpicture}
\\
{\scriptsize \textbf{(b)} Regions in the potential space.}
\end{center}
\end{minipage}
\caption{Regions in the potential space bounded by hyperplanes induced by breakpoints of the marginal edge cost functions (dashed lines). The optimal potentials inside the regions form line segments (thick lines).}
\label{fig:potentialspace}
\end{figure}

Assume that the marginal cost functions are piecewise linear, i.e., every marginal cost function has $\bar{t}_e \in \N$ linear function parts and $\bar{t}_e + 1$ breakpoints $\tau_{e, t_e}, t_e \in \{1, \dotsc, \bar{t}_e + 1\}$, including the artificial breakpoints $\tau_{e,1} = - \infty$ and $\tau_{e, \bar{t}_e +1} = \infty$. Denote by $\sigma_{e, t_e} := \mcost (\tau_{e, t_e})$ the value at the respective breakpoint. For each function part~$t_e \in \{1, \dotsc, \bar{t}_e\}$, there are coefficients $\alpha_{e, t_e} > 0$ and $\beta_{e, t_e} \in \R$ such that $\mcost (x) = \alpha_{e, t_e} x + \beta_{e, t_e}$ for all $x$ between the breakpoints of the respective function part. The inverses of the marginal cost functions exist since $\alpha_e > 0$ and have the explicit form $\mcost^{-1} (y) = c_{e, t_e} y + d_{e, t_e}$ where $c_{e, t_e} = \frac{1}{a_{e, t_e}}$ and $d_{e, t_e} = \frac{b_{e, t_e}}{a_{e, t_e}}$. Denote by $\vec{f}^{-1}$ the vector of all inverse marginal cost functions. Then $\vec{f}^{-1}$ is also a piecewise linear function, where every linear part can be expressed as
\[
\vec{f}^{-1} (\vec{y}) = \vec{C}_{\vec{t}} \vec{y} - \vec{d}_{\vec{t}},
\]
where $\vec{C}_{\vec{t}} := \diag ( c_{e_1, t_{e_1}}, \dotsc, c_{e_m, t_{e_m}} )$ and $\vec{d}_{\vec{t}} := (d_{e, t_e})_{e \in E}$ and the vector $\vec{t} = (t_e)_{e \in E}$ encodes the respective function part.
For every potential $\vec{\phi} \in \Pi$ from the potential space, the flow $\vec{x} := \vec{f}^{-1} ( \vec{\Gamma} \vec{\phi})$ satisfies, together with the potential~$\vec{\phi}$, the optimality conditions~\eqref{eq:kkt:undirected} for the undirected minimum-cost flow problem by definition.
The function $\vec{\phi} \mapsto \vec{f}^{-1} (\vec{\Gamma} \vec{\phi})$ is piecewise linear and has the explicit form $\vec{f}^{-1} (\vec{\Gamma} \vec{\phi}) = \vec{C}_{\vec{t}} \vec{\Gamma}^{\top} \vec{\phi} - \vec{d}_{\vec{t}}$ on the polytopes
\[
R_{\vec{t}} := \{ \vec{\phi} \in \Pi \mid \sigma_{e, t_e} \leq \phi_w - \phi_v \leq \sigma_{e, t_e + 1} \text{ for all } e = (v,w) \in E \}
\]
that we refer to as \emph{regions}. Figure~\ref{fig:potentialspace} shows these regions in the potential space for a concrete example. A potential $\vec{\pi} \in \Pi$ therefore is an optimal potential of the minimum-cost flow $\vec{x}^* = \vec{C}_{\vec{t}} \vec{\Gamma}^{\top} \vec{\pi} - \vec{d}_{\vec{t}}$ if and only if $\vec{x}^*$ satisfies the demands $\vec b^0 + \lambda \vec b$, i.e., if and only if $\vec{\Gamma} \vec{x}^*  = \vec{b}^0 + \lambda \vec{b}$ which is equivalent to
\begin{equation} \label{eq:laplace}
\vec{L}_{\vec{t}} \vec{\pi} - \tilde{\vec{d}}_{\vec{t}} = \vec{b}^0 + \lambda \vec{b},
\end{equation}
where $\vec{L}_{\vec{t}} := \vec{\Gamma} \vec{C}_{\vec{t}} \vec{\Gamma}$ and $\tilde{\vec{d}}_{\vec{t}} = \vec{\Gamma} \vec{d}_{\vec{t}}$. Since the Laplacian matrix $\vec{L}_{\vec{t}}$ has rank $n-1$, \eqref{eq:laplace} has a unique solution $\vec{\pi} \in \Pi$ that can be expressed explicitly as
\[
\vec{\pi} (\lambda) = \vec{\pi}_{\vec{t}} + \lambda \Delta \vec{\pi}_{\vec{t}},
\]
where $\Delta \vec{\pi}_{\vec{t}} := \vec{L}^*_{\vec{t}} \vec{b}$ and $\vec{\pi}_{\vec{t}} := \vec{L}^*_{\vec{t}} (\vec{b}^0 + \tilde{\vec{d}}_{\vec{t}})$. If the solution $\vec{\pi} (\lambda)$ is inside the region $R_{\vec{t}}$, then $\vec{\pi}(\lambda)$ is an optimal potential for the parameter~$\lambda$. Therefore, the set of all optimal potentials in the region $R_{\vec{t}}$ is
\[
\Pi_{\vec{t}} = \{ \vec{\pi}_{\vec{t}} + \lambda \Delta \vec{\pi}_{\vec{t}} \mid \lambda \geq 0 \text{ and } \vec{\pi}_{\vec{t}} + \lambda \Delta \vec{\pi}_{\vec{t}} \in R_{\vec{t}} \}
\]
and, thus, it is a (possibly empty) line segment. The following theorem summarizes these observations.
For more details and a rigorous proof, see~\cite{KlimmWarode2021}.

\begin{theorem}
For every region $R_{\vec{t}}$ there are numbers $\lambda^{\min}_{\vec{t}}, \lambda^{\max}_{\vec{t}} \in \R \cup \{ \infty \}$ such $\vec{\pi} \in \Pi_{\vec{t}} =  \{ \vec{\pi}_{\vec{t}} + \lambda \, \Delta \vec{\pi}_{\vec{t}} \, \vert \, \lambda^{\min}_{\vec{t}} \leq \lambda \leq \lambda^{\max}_{\vec{t}} \}$.
A potential vector $\vec{\pi} \in R_{\vec{t}}$ with $\pi_{v_1} = 0$ is the potential of an optimal solution of~\eqref{eq:undirected_mincost_flow} for some $\lambda \geq 0$ if and only if $\vec{\pi} \in \Pi_{\vec{t}}$.
\end{theorem}

Figure~\ref{fig:potentialspace} shows the linear line segments of optimal potentials in the regions for a concrete example. The (non-empty) line segments of neighboring regions intersect in the boundary hyperplanes separating the regions. Every boundary corresponds to a breakpoint of some marginal cost function; the crossing of a boundary corresponds to moving from one function part of a marginal cost function to the next. Given a region with non-empty line segment $\Pi_{\vec{t}}$, the boundary inducing the maximal parameter value $\lambda^{\max}_{\vec{t}}$ leads to a neighboring region that also has a non-empty line segment. We can find such boundaries in $\mathcal{O}(n^2)$-time. If the boundary, and therefore the neighboring region, is unique, the (pseudo-)inverse of the Laplacian matrix $\vec{L}^*_{\vec{t}}$ of the neighboring region can also be obtained in $\mathcal{O}(n^2)$-time via a simple rank-1-update formula~\cite[Thm.~7]{KlimmWarode2021}. If the boundary is non-unique, i.e., if the line segment of optimal potentials ends in the intersection of multiple boundaries, a unique neighboring region can either be obtained using a lexicographic rule or by solving a quadratic program.

Overall, this leads to the following procedure for the computation of the optimal potential function $\lambda \mapsto \vec{\pi} (\lambda)$ (and therefore also the minimum-cost flow function $\lambda \mapsto \vec{x}^* (\lambda)$.
\begin{enumerate}
\item
	Compute an initial solution $\vec{x}^* (0)$ with an optimal potential $\vec{\pi} (0)$ for the demand $\vec{b}^0$. In some cases this initial solution is trivial. For example, if $\vec{b}^0 = \vec{0}$ and the marginal cost are homogeneous, meaning that $\mcost (0) = 0$, the zero flow and zero potential are the optimal solution for $\lambda = 0$. Otherwise, the initial solution can be obtained by solving the minimum-cost flow problem for the fixed demand $\vec{b}^0$ in strongly polynomial time, for instance with the algorithm by Végh~\cite{vegh2016strongly}.
\item
	Find the initial region $R_{\vec{t}}$ for the initial potential and compute the pseudo-inverse of the Laplacian $\vec{L}^*_{\vec{t}}$.
\item
	Compute the direction vector $\Delta \vec{\pi}_{\vec{t}}$, the offset vector $\vec{\pi}_{\vec{t}}$, and the maximal parameter value $\lambda^{\max}_{\vec{t}}$ for this region.
\item
	As long as $\lambda^{\max}_{\vec{t}} < \infty$, find the neighboring region $R_{\vec{t}'}$ with non-empty line segment $\Pi_{\vec{t}'}$ and continue with step~3 for $\vec{t} = \vec{t}'$.
\end{enumerate}
This procedure computes the complete piecewise linear function $\lambda \mapsto \vec{\pi} (\lambda)$. Every iteration yields one function part defined by the offsets $\vec{\pi}_{\vec{t}}$ and directions $\Delta \vec{\pi}_{\vec{t}}$, and the minimal and maximal parameter values $\lambda^{\min}_{\vec{t}}$ and $\lambda^{\max}_{\vec{t}}$. With the function $\vec{\phi} \mapsto \vec{f}^{-1} (\vec{\Gamma} \vec{\phi})$, we can also compute the offsets and directions of the flow function $\lambda \mapsto \vec{x}^* (\lambda)$.

\subsubsection{Directed parametric minimum-cost flows}

The algorithm described in the previous subsection works only for undirected minimum-cost flow problems, i.e., problems without any constraint on the flow. Assume, we are given lower capacities $\vec{l} = (l_{e})_{e \in E}$ and upper capacities $\vec{u} = (u_e)_{e \in E}$ for the edge flows and consider the minimum-cost flow problem
\[
\min \quad \sum_{e \in E} F_e(x_e) \quad
\text{s.t.} \quad \vec \Gamma \vec x = \vec{b}^0 + \lambda \vec b, \quad
\vec{l} \leq \vec x \leq \vec{u},
\]
where the edge cost are the integral of (piecewise linear) marginal costs, i.e., $\ecost (x) = \int_{l_e}^x \mcost (s) ds$. In the special case of $l_e = 0$ and $u_e = \infty$ for all $e \in E$, this problem is exactly the parametric minimum-cost flow problem from~\eqref{eq:mincost_flow}. It is not hard to see that $\vec{x}^*$ is an optimal solution to the above problem if and only if $\vec{x}^*$ is an optimal solution to the undirected problem
\[
\min \quad \sum_{e \in E} \tilde{F}_e(x_e) \quad
\text{s.t.} \quad \vec \Gamma \vec x = \vec{b}^0 + \lambda \vec b
\]
with the special cost functions $\tilde{F}_{e} (x) := \int_{l_e}^x \tilde{f}_e (s) ds$ with $\tilde{f}_e (s) = - \infty$ if $s < l_e$, $\tilde{f}_e (s) = \mcost (s)$ if $l_e \leq s \leq u_e$, and $\tilde{f}_e (s) = \infty$ if $s > u_e$, where we interpret the integral over the infinite function parts as infinite. Thus, we can reformulate every minimum-cost flow problem with flow bounds (and, hence, every directed problem) as an undirected problem with infinite costs.
Although the marginal costs in the undirected problem take infinite values, we can still define a meaningful inverse function $\mcost^{-1} (y) = c_{e, t_e} y + d_{e, t_e}$ by setting $c_{e, t_e} = 0$ for the function parts where $\mcost$ is infinite. This definition ensures that for every potential $\vec{\phi}$, the flow $\vec{x} := \vec{f}^{-1} (\vec{\Gamma} \vec{\phi})$, we have $\mcost(x_e) = \pi_w - \pi_v$ for all $e = (v,w)$ with $l_e \leq x_e \leq u_e$, $\mcost(x_e) \geq \phi_w - \phi_v$ for all $e = (v,w)$ with $x_e < l_e$ and $\mcost(x_e) \leq \phi_w - \phi_v$ for all $e = (v, w)$ with $x_e > u_e$. The latter conditions are exactly the KKT conditions for the bounded problem, hence, the function $\vec{\phi} \mapsto \vec{f}^{-1} (\vec{\Gamma} \vec{\phi})$ maps every potential to a flow that satisfies the optimality conditions.

Therefore, we can use the exact same algorithm as for the undirected problem to solve the parametric problem. However, since we now admit that $c_{e, t_e}$ can be zero, it can happen that the Laplacian $\vec{L}_{\vec{t}}$ matrix of some regions has rank smaller than $n - 1$. In such regions, the pseudo-inverse  of the Laplacian is no longer unique, leading to non-unique direction and offset vectors. We therefore call these regions \emph{ambiguous}. In \cite[Section~4.2]{KlimmWarode2021}, we show that for every ambiguous region, there is a (unique) way to skip the region and proceed in another, non-ambiguous region without changing the flow.

\subsection{Minimum-cost Flows With General Convex Cost} \label{sec:generalcost}

In this subsection, we develop two algorithms for the parametric computation of minimum-cost flows for general convex cost. Formally, we consider the parametric problem
\begin{equation} \label{eq:apx:mcflow:problem}
\min \quad \sum_{e \in E} F_e(x_e) \quad
\text{s.t.} \quad \vec \Gamma \vec x = \vec{b}^0 + \lambda \vec b, \quad
\vec{l} \leq \vec x \leq \vec{u},
\end{equation}
for general flow bounds $l_e, u_e \in \R \cup \{- \infty, \infty\}$, i.e., we consider the directed as well as the undirected case, and edge cost $F_e (x) := \int_{l_e}^x \mcost(s) ds$ that are the integral over a strictly increasing, differentiable marginal cost function $\mcost$ with $\mcost (s) \geq 0$ for all $s > 0$ and $\mcost (s) \leq 0$ for all $s < 0$.

Since for general, non-linear marginal cost functions $\mcost$ the flows and solution can be irrational, we can only aim for the approximate solutions. In order to obtain meaningful approximation guarantees, we restrict the parametric computation to all $\lambda \in [0, \lambda^{\max}]$ for some $\lambda^{\max} > 0$.
For given $\alpha > 1$ and $\beta \geq 0$, we say a function $\tilde{\vec{x}} \colon [0, \lambda^{\max}] \to \R^m$ is an \emph{$(\alpha, \beta)$-approximate minimum-cost flow function} on $[0, \lambda^{\max}]$ if for every $\lambda \in [ 0, \lambda^{\max} ]$, the vector $\tilde{\vec{x}} (\lambda)$ is a feasible flow for the demand $\vec{b}^0 + \lambda \vec{b}$ and
\[
\tilde{C} (\lambda) \leq \alpha C(\lambda) + \beta
\qquad
\text{for all } \lambda \in [0, \lambda^{\max}],
\]
where $\tilde{C} (\lambda) := \sum_{e \in E} F_e (\tilde{x}_e )$ denotes the cost of the flow $\tilde{\vec x} (\lambda)$. An approximate minimum-cost flow function maps every parameter $\lambda$ to a feasible flow whose cost do not exceed the optimal cost by an absolute factor of $\beta$ and a relative factor of $\alpha$. 
We proceed by developing two methods that compute an approximate minimum-cost flow function.

\subsubsection{Marginal cost approximation}

The first method for the approximate computation of parametric minimum-cost flows is based on the approximation of the marginal cost of the given parametric problem. Instead of computing a solution to the actual problem, we consider a new instance with marginal cost $\tilde{f}_e$ that are a piecewise linear interpolation of the original marginal costs $\mcost$. This new instance therefore has piecewise quadratic edge cost and can be solved exactly with the algorithm from the previous section. Since this method is based on an approximation of the marginal cost, we refer to it as \emph{marginal cost approximation}~(MCA). 

The idea behind the marginal cost approximation is that we can choose a small mesh size of the interpolation in order to obtain a good approximation of the marginal cost. Intuitively, the minimum-cost flow functions of two instances with similar marginal cost functions should not differ too much. The following lemma proves, that this is indeed the case. 

We denote by $x^{\max} := \max_{\lambda \in [0, \lambda^{\max}]} \frac{1}{2} \sum_{v \in V} |b^0_v + \lambda b_v|$ the maximal total inflow in the network over all parameter values. Since we assume that the marginal cost functions $\mcost$ are homogeneous, i.e., $\mcost (s) \leq 0$ for $s < 0$ and $\mcost (s) \geq 0$ for $s > 0$, all optimal flows must be cycle-free and therefore $x^{\max}$ is a trivial bound for every edge flow $x_e (\lambda)$ for all edges $e \in E$ and all $\lambda \in [0, \lambda^{\max}]$.

\begin{lemma} \label{lem:mca}
Let $\alpha > 1$, $\beta \geq 0$, and $\lambda^{\max} > 0$. Given two families of marginal cost functions $(\mcost)_{e \in E}$ and $(\tilde\mcost)_{e \in E}$. Let $\lambda \mapsto \vec{x}^* (\lambda)$ be the minimum-cost flow function solving the problem~\eqref{eq:apx:mcflow:problem} for edge cost $\ecost (x) := \int_{l_e}^x \mcost (s) ds$ and $\lambda \mapsto \tilde{\vec{x}} (\lambda)$ be the minimum-cost flow function solving the problem~\eqref{eq:apx:mcflow:problem} for the edge cost $\tilde\ecost (x) := \int_{l_e}^x \tilde\mcost (s) ds$.
If
\begin{equation} \label{eq:lem:mca:1}
\big| \tilde\mcost( x ) - \mcost (x) \big| \leq
\frac{\alpha - 1}{1 + \alpha} \big|\mcost (x)\big| + \frac{\beta}{(1+\alpha) m x^{\max}}
\end{equation}
for all $- x^{\max} \leq x \leq x^{\max}$ and $e \in E$ 
or 
if $|\tilde{\mcost} (x)| \geq | \mcost (x) |$ and
\begin{equation} \label{eq:lem:mca:2}
\big| \tilde\mcost( x ) - \mcost (x) \big| \leq
(\alpha - 1) \big| \mcost (x) \big| + \frac{\beta}{m x^{\max}}
\end{equation}
for all $- x^{\max} \leq x \leq x^{\max}$ and $e \in E$, then $\lambda \mapsto \tilde{\vec{x}} (\lambda)$ is an $(\alpha, \beta)$-approximate minimum-cost flow function on~$[0, \lambda^{\max}]$ for the minimum-cost flow problem with edge cost $\ecost$.
\end{lemma}
\begin{proof}
Assume that the marginal cost satisfy~\eqref{eq:lem:mca:1}. Then, for every edge $e \in E$ and every $-x^{\max} \leq s \leq x^{\max}$, we have
\[
- \frac{\alpha - 1}{1 + \alpha} |\mcost (s)| - \frac{\beta}{(1+\alpha) m x^{\max}} 
\leq 
\tilde\mcost( s ) - \mcost (s) 
\leq
\frac{\alpha - 1}{1 + \alpha} |\mcost (s)| + \frac{\beta}{(1+\alpha) m x^{\max}}
.
\]
Since we assume $\mcost (s) \geq 0$ if $s > 0$ and $\mcost (s) \leq 0$ if $s < 0$, this is equivalent to 
\begin{align*}
\frac{2}{1 + \alpha} \mcost (s) - \frac{\beta}{(1 + \alpha) m x^{\max}} \leq \tilde{\mcost} (s) &\leq \frac{2 \alpha}{(1+\alpha)} \mcost (s) + \frac{\beta}{(1 + \alpha) m x^{\max}}
\text{ if } s \geq 0 \\
\frac{2 \alpha}{1 + \alpha} \mcost (s) - \frac{\beta}{(1 + \alpha) m x^{\max}} \leq \tilde{\mcost} (s) &\leq \frac{2}{(1+\alpha)} \mcost (s) + \frac{\beta}{(1 + \alpha) m x^{\max}}
\text{ if } s < 0
.
\end{align*}
Taking the integral from $l_e$ to $x_e$ over the above inequalities, we obtain 
\begin{equation} \label{eq:prf:lem:mca:1}
\frac{2}{1+\alpha} \ecost (x) - \frac{\beta}{(1 + \alpha) m}
\leq 
\tilde{\ecost} (x) \leq 
\frac{2 \alpha}{1 + \alpha} \ecost (x) + \frac{\beta}{(1+ \alpha) m}
\end{equation}
for all $-x^{\max} \leq x \leq x^{\max}$.
We note that $\sum_{e \in E} \tilde\ecost (\tilde{x}_e (\lambda)) \leq \sum_{e \in E} \tilde{\ecost} (x^*_e (\lambda))$ since $\tilde{\vec{x}} (\lambda)$ is the minimum-cost flow for parameter~$\lambda$ with respect to the edge cost~$\tilde{\ecost}$. Further, since $x^{\max}$ is a bound on the maximal flow on any edge, we can apply~\eqref{eq:prf:lem:mca:1} for $x = x^*_e (\lambda)$ and $x=\tilde{x}_e (\lambda)$ and obtain
\begin{align*}
\tilde{C} (\lambda) 
&= \sum_{e \in E} \ecost ( \tilde{x}_e (\lambda))
\stackrel{\mathclap{\eqref{eq:prf:lem:mca:1}}}{\leq}
\sum_{e \in E} \frac{1+\alpha}{2} \bigg( \tilde{\ecost} (\tilde{x}_e (\lambda)) + \frac{\beta}{(1+\alpha) m} \bigg) \\
&\leq \sum_{e \in E} \frac{1+\alpha}{2} \bigg( \tilde{\ecost} (x^*_e( \lambda)) + \frac{\beta}{(1+ \alpha) m} \bigg) \\
&\stackrel{\mathclap{\eqref{eq:prf:lem:mca:1}}}{\leq}
\sum_{e \in E} \frac{1+\alpha}{2} \bigg( \frac{2 \alpha}{1+ \alpha} \ecost (x^*_e( \lambda)) + \frac{2 \beta}{(1+ \alpha) m}  \bigg) \\
&= \alpha \sum_{e \in E} \ecost ( x^*_e (\lambda)) + \sum_{e \in E} \frac{\beta}{m}
= \alpha C(\lambda) + \beta
.
\end{align*}
Thus, $\lambda \mapsto \tilde{\vec{x}} (\lambda)$ is an $(\alpha, \beta)$-approximate minimum-cost flow function.

If we assume that $\tilde{\mcost} (x)| \geq | \mcost (x) |$, then $\tilde\ecost (x) \geq \ecost (x)$. Similar to the first case, it can be shown that~\eqref{eq:lem:mca:2} implies $\tilde\ecost (x) \leq \alpha \ecost (x) + \frac{\beta}{m}$. Overall, we obtain
\[
\tilde{C} (\lambda) = \sum_{e \in E} \ecost (\tilde{x}_e (\lambda)) 
\leq \sum_{e \in E} \tilde\ecost (\tilde{x}_e (\lambda)) 
\leq \sum_{e \in E} \bigg( \alpha \ecost (x) + \frac{\beta}{m} \bigg) = \alpha C (\lambda) + \beta
\]
and the claim follows.
\end{proof}

Lemma~\ref{lem:mca} shows that we can obtain an approximate minimum-cost flow if we solve a different problem with marginal cost functions that differ only slightly from the original marginal cost functions. By using a linear spline interpolation of the marginal cost functions, we obtain a second instance with piecewise linear marginal cost functions that we can solve with the method from the previous subsection. Further, by choosing a sufficiently small interpolation mesh, we can interpolate the marginal costs with arbitrarily small error and apply Lemma~\ref{lem:mca}.
Formally, let $[a,b]$ be some interval and $a = x_0 < x_1 < \dots < x_K = b$ be a family of breakpoints that we also refer to as mesh. Then, given some function $f \colon [a,b] \to \R$, we call the piecewise linear function defined as
\[
s_f \colon [a, b] \to \R, x \mapsto s_f (x) := f(x_i) + \frac{x - x_i}{x_{i+1} - x_i} \big( f (x_{i+1}) - f(x_i) \big) 
\]
for $x \in [x_i, x_{i+1}]$ the \emph{linear spline of $f$ with mesh $(x_i)_{i=1, \dotsc, K}$}. We denote by $\delta_i := x_{i+1} - x_i$ the distance of two adjacent mesh points and refer to $\delta_i$ as the $i$-th \emph{step size} and call $\max_i \delta_i$ the \emph{mesh size} of the interpolation. If the function~$f$ is twice differentiable, then basic calculus (see, e.g., the textbook by de~Boor~\cite{de1978practical} for a reference) yields that
\begin{equation} \label{eq:linear_spline}
| f(x) - s_f (x)| \leq \frac{1}{8} \delta^2_i \, \max_{\xi \in [x_i, x_{i+1}]} |f''(\xi)|
\end{equation}
for every $x \in [x_i, x_{i+1}]$.
Combining this with Lemma~\ref{lem:mca} yields the following result.

\begin{theorem} \label{thm:mca}
Given a parametric minimum-cost flow problem with marginal cost functions $\mcost$ that are differentiable three times, let $\alpha > 1$, $\beta \geq 0$, and $\lambda^{\max} > 0$. Further, for every edge, let $s_{f_e}$ be a linear spline of the marginal cost with the some mesh $-x^{\max} = x_1 < \dotsc < x_K = x^{\max}$. If 
\begin{enumerate}[(i)]
\item
	the step sizes $\delta_i = x_{i+1} - x_i$ satisfy
	\[
	\delta_i \sqrt{\max_{\xi \in [x_i, x_{i+1} + \delta_i]} |\mcost''(\xi)|}
	\leq
	2 \sqrt{2} 
	\sqrt{
	\frac{\alpha - 1}{1 + \alpha} |\mcost (x_i)| 
	+
	\frac{\beta}{(1+\alpha) m x^{\max}}
	}
	,
	\]
	or
\item
	for all edges $e \in E$, $\mcost'' (x)$ is non-decreasing and $\mcost'' (x) \geq 0$ for $x \geq 0$ and $\mcost'' (x)$ is non-increasing and $\mcost'' (x) \leq 0$ for $x < 0$ and the step sizes $\delta_i = x_{i+1} - x_i$ satisfy
	\[
	\delta_i \, B \leq 2 \sqrt{2} \sqrt{ (\alpha - 1) | \mcost (x_i) | + \frac{\beta}{m x^{\max}} },
	\]
	where $B := \max \{ \sqrt{|\mcost'' (x_i)|, |\mcost'' (x_i + \delta_i)|} \}$,
\end{enumerate}
then the minimum-cost flow function $\lambda \mapsto \tilde{\vec{x}} (\lambda)$ of the instance with marginal cost functions $s_{\mcost}$ is an $(\alpha, \beta)$-approximate minimum-cost flow function on $[0, \lambda^{\max}]$ for the original minimum-cost flow problem.
\end{theorem}
\begin{proof}
With~\eqref{eq:linear_spline}, we can use \emph{(i)} in order to obtain
\[
|\mcost (x) - s_{\mcost} (x)| 
\leq 
\frac{1}{8} \delta_i^2 \max_{\xi \in [x_i, x_{i+1} + \delta_i]} |\mcost''(\xi)|
\stackrel{\emph{(i)}}{\leq}
\frac{\alpha-1}{1+\alpha} |\mcost(x)| + \frac{\beta}{(1+\alpha) m x^{\max}}
\]
and, hence, Lemma~\ref{lem:mca} proves the claim.

If we assume \emph{(ii)} is satisfied, then the marginal cost functions are convex for $x > 0$ and concave for $x < 0$. Thus, the linear spline $s_{\mcost}$ is always smaller than the function $\mcost$ for negative values and grater than the function $\mcost$ for positive values. Thus, $|s_{\mcost} (x)| \geq |\mcost (x)$ for all $x \in \R$. Additionally, the monotonicity of $\mcost''$, we get that $\max_{\xi \in [x_i, x_{i+1}]} |\mcost'' (\xi)| = \max \{ |\mcost''(x_i)|, | \mcost''(x_i + \delta_i)| \} = B^2$. Hence, with~\eqref{eq:linear_spline} we obtain
\[
|\mcost (x) - s_{\mcost} (x)| 
\leq 
\frac{1}{8} \delta_i^2 B^2 
\stackrel{\emph{(ii)}}{\leq}
(\alpha-1) |\mcost (x)| + \frac{\beta}{m x^{\max}}
\]
and the claim follows from Lemma~\ref{lem:mca}.
\end{proof}

\subsubsection{Minimum-cost flow interpolation}

The second method for the approximate computation of parametric minimum-cost flows is based on the following, intuitive idea. If we are given two minimum-cost flows $\vec{x}^1$ and $\vec{x}^2$ for two different parameter values $\lambda_1 < \lambda_2$, then every convex combinations $\frac{\lambda - \lambda_2}{\lambda_2 - \lambda_1} \vec{x}^1 + \frac{\lambda - \lambda_1}{\lambda_2 - \lambda_1} \vec{x}^2$ is a feasible flow for the parameter $ \lambda \in [\lambda_1, \lambda_2]$. Further, if the breakpoints $\lambda_1$ and $\lambda_2$ are not too far apart, the convex combinations should also be approximate minimum-cost flows. Therefore, our second approach that we refer to as \emph{minimum-cost flow interpolation}~(MCFI) works as follows. First, we choose fixed parameter values $0 = \lambda_1 < \lambda_2 < \dots < \lambda_K = \lambda^{\max}$ with \emph{step sizes} $\delta_i = \lambda_{i+1} - \lambda_i$. Then we compute minimum-cost flows $\vec{x}^i$ for every breakpoint $\lambda_i$. Finally, we compute a linear interpolation $\lambda \mapsto \tilde{\vec{x}}$ of the points $(\lambda_i, \vec{x}^i)$ and use this as an approximate minimum-cost flow function. 
The aim of this subsection is to find bounds on the step sizes that guarantee that the output of this method is indeed an $(\alpha, \beta)$-approximate minimum-cost flow function.

Since in general it is not possible to compute an exact solution even for fixed parameter values, we assume that we are given some oracle $\mathcal{X}$ that returns a minimum-cost flow for every fixed parameter value~$\lambda$ with $\fwprec$~precision, i.e., $\mathcal{X} (\lambda)$ is a feasible flow for the demand $\vec{b}^0 + \lambda \vec{b}$ such that $C(\mathcal{X} (\lambda)) \leq (1 + \epsilon) C(\lambda)$, where $C(\lambda)$ are the cost of an optimal solution.

Given a family of fixed parameter values $0 = \lambda_1 < \lambda_2 < \dots < \lambda_K = \lambda^{\max}$, we define the function
\begin{equation} \label{eq:mcfi:function}
\tilde{\vec{x}} \colon [0, \lambda^{\max}], \lambda \mapsto 
\frac{\lambda - \lambda_{i+1}}{\lambda_{i+1} - \lambda_i} \mathcal{X}( \lambda_{i} ) + \frac{\lambda - \lambda_i}{\lambda_{i+1} - \lambda_i} \mathcal{X}( \lambda_{i+1} )
\quad
\text{if } \lambda \in [\lambda_i, \lambda_{i+1}]
.
\end{equation}

\begin{lemma} \label{lem:mcfi}
Let $\alpha > 1$, $\beta \geq 0$, and $0 < \epsilon < \alpha - 1$. Let $0 = \lambda_1 < \lambda_2 < \dots < \lambda_K = \lambda^{\max}$ be the breakpoints of the function $\lambda \mapsto \tilde{\vec{x}} (\lambda)$ defined in~\eqref{eq:mcfi:function}. If for every interval $[\lambda_i, \lambda_{i+1}]$ one of the following conditions holds true, then $\lambda \mapsto \tilde{\vec{x}} (\lambda)$ is an $(\alpha, \beta)$-approximate minimum-cost flow function.
\begin{enumerate}[(i)]
\item
	The objective function $C(\lambda)$ is non-decreasing on $[\lambda_i, \lambda_{i+1}]$ and the step size $\delta_i := \lambda_{i+1} - \lambda_i$ satisfies
	\[
	\delta_i \frac{d}{d \lambda} C(\lambda_i + \delta_i) \leq \frac{\alpha - 1 - \epsilon}{1 + \epsilon} C(\lambda_i) + \frac{\beta}{1+ \epsilon}
	.
	\]
\item
	The objective function $C(\lambda)$ is non-increasing on $[\lambda_i, \lambda_{i+1}]$ and the step size $\delta_i := \lambda_{i+1} - \lambda_i$ satisfies
	\[
	\delta_i \frac{d}{d \lambda} C(\lambda_i) \geq - \frac{\alpha - 1 - \epsilon}{1 + \epsilon} C(\lambda_i) - \frac{\beta}{1+ \epsilon}
	\]
\item
	For every $e \in E$, the marginal cost function $\mcost$ is differentiable and $\mcost' (x) > 0$ for all $x$. Further, the support does not change between $\lambda_i$ and $\lambda_{i+1}$, i.e., $S (\lambda) = S(\lambda_i)$ for all $\lambda \in [\lambda_i, \lambda_{i+1}]$ and the step size $\delta_i := \lambda_{i+1} - \lambda_i$ satisfies
	\[
	\delta_i
	\max_{\lambda \in [\lambda_i, \lambda_i + \delta_i] }
	\sqrt{\vec{b}^{\top} \vec{L}^*_{\lambda} \vec{b}}
	\leq
	2 \sqrt{2}
	\sqrt{\frac{\alpha - 1 - \epsilon}{1 + \epsilon} C(\lambda_i) + \frac{\beta}{1  + \epsilon}}
	\]
\end{enumerate}
\end{lemma}
\begin{proof}
By Theorem~\ref{thm:derivative_objective}, the function $\lambda \mapsto C(\lambda)$ is convex. Thus, $\lambda \mapsto \frac{d}{d \lambda} C (\lambda)$ is non-decreasing. For $\lambda \in [\lambda_i, \lambda_{i+1}]$, we obtain with the mean value theorem that there exists $\xi \in [\lambda_i, \lambda_{i+1}]$ such that
\begin{equation} \label{eq:prf:lem:mcfi:1}
\frac{d}{d \lambda} C(\lambda_i) \leq
\frac{C(\lambda_{i+1}) - C(\lambda_i)}{\lambda_{i+1} - \lambda_i}
=
\frac{d}{d \lambda} C(\xi)
\leq
\frac{d}{d \lambda} C(\lambda_{i+1})
.
\end{equation}
Since the mapping $\vec{x} \mapsto C(\vec{x})$ (i.e., the function mapping a flow to its cost) is also convex, we obtain with condition~\emph{(i)} that
\begin{align*}
\tilde{C} (\lambda)
&\leq 
\frac{\lambda - \lambda_i}{\lambda_{i+1} - \lambda_i} C( \mathcal{X} (\lambda_{i+1})) + \frac{\lambda_{i+1} - \lambda}{\lambda_{i+1} - \lambda_i} C(\mathcal{X} (\lambda_i)) \\
&\leq
(1+ \epsilon)
\bigg(
\frac{\lambda - \lambda_i}{\lambda_{i+1} - \lambda_i} \big( C(\lambda_{i+1}) - C(\lambda_i) \big) + C(\lambda_i) \bigg)
\\
&\stackrel{\mathclap{\eqref{eq:prf:lem:mcfi:1}}}{\leq}
(1+ \epsilon)
\bigg(
\delta_i \frac{d}{d \lambda} C(\lambda_{i+1}) + C(\lambda_i) \bigg)
\bigg)
\stackrel{\emph{(i)}}{\leq}
\alpha C(\lambda_i) + \beta \leq \alpha C(\lambda) + \beta
,
\end{align*}
where we used that $\lambda \mapsto C(\lambda)$ is non-decreasing on $[\lambda_i, \lambda_{i+1}]$. This proves the sufficiency of condition \emph{(i)}. For the second condition, we observe that
\begin{align*}
\tilde{C} (\lambda)
&\leq 
\frac{\lambda - \lambda_i}{\lambda_{i+1} - \lambda_i} C( \mathcal{X} (\lambda_{i+1})) + \frac{\lambda_{i+1} - \lambda}{\lambda_{i+1} - \lambda_i} C(\mathcal{X} (\lambda_i)) \\
&\leq
(1+ \epsilon)
\bigg(
\frac{\lambda - \lambda_i}{\lambda_{i+1} - \lambda_i} \big( C(\lambda_i) - C(\lambda_{i+1}) \big) + C(\lambda_{i+1}) \bigg)
\\
&\stackrel{\mathclap{\eqref{eq:prf:lem:mcfi:1}}}{\leq}
(1+ \epsilon)
\bigg(
-\delta_i \frac{d}{d \lambda} C(\lambda_{i}) + C(\lambda_{i+1}) \bigg)
\bigg)
\\
&\stackrel{\mathclap{\emph{(ii)}}}{\leq}
(1+ \epsilon)
\bigg(
\frac{\alpha - 1 - \epsilon}{1 + \epsilon} C (\lambda_i)
+ \frac{\beta}{1 + \epsilon} + C(\lambda_{i+1})
\bigg) \\
&=
\alpha C(\lambda_{i+1}) + \beta \leq \alpha C(\lambda) + \beta,
\end{align*}
where we used that $\lambda \mapsto C(\lambda)$ is non-increasing on $[\lambda_i, \lambda_{i+1}]$. This proves that condition~\emph{(ii)} is sufficient. For the third condition, we define the function 
\[
g(\lambda) := \frac{\lambda - \lambda_i}{\lambda_{i+1} - \lambda_i} C(\lambda_{i+1}) + \frac{\lambda_{i+1} - \lambda}{\lambda_{i+1} - \lambda_i} C(\lambda_i)
.
\]
This function is a linear spline interpolating the function $C(\vec{x})$ on the interval $[\lambda_i, \lambda_{i+1}]$. Since we assume that the support is unchanged on this interval, the function $\lambda \mapsto C(\lambda)$ is twice differentiable on $[\lambda_i, \lambda_{i+1}]$ by Theorem~\ref{thm:derivative_objective}. With~\eqref{eq:linear_spline} we obtain
\begin{equation} \label{eq:prf:lem:mcfi:2}
g(\lambda) - C(\lambda) \leq \frac{1}{8} \delta_i^2 \max_{\lambda \in [\lambda_i, \lambda_{i+1}]} \frac{d^2}{d^2 \lambda} C(\lambda) = \frac{1}{8} \delta_i^2 \max_{\lambda \in [\lambda_i, \lambda_{i+1}]} \vec{b}^{\top} \vec{L}_{\lambda} \vec{b}
.
\end{equation}
Therefore, we get
\begin{align*}
\tilde{C} (\lambda)
&\leq
\frac{\lambda - \lambda_i}{\lambda_{i+1} - \lambda_i} C( \mathcal{X} (\lambda_{i+1})) + \frac{\lambda_{i+1} - \lambda}{\lambda_{i+1} - \lambda_i} C(\mathcal{X} (\lambda_i)) \\
&\leq
(1 + \epsilon) \big( g(\lambda) - C(\lambda) \big) + (1 + \epsilon) C(\lambda) \\
&\stackrel{\mathclap{\eqref{eq:prf:lem:mcfi:2}}}{\leq}
\frac{1+\epsilon}{8} \delta_i^2  \max_{\lambda \in [\lambda_i, \lambda_{i+1}]} \vec{b}^{\top} \vec{L}_{\lambda} \vec{b} 
+ (1 + \epsilon) C(\lambda) \\
&\stackrel{\mathclap{\emph{(iii)}}}{\leq}
( \alpha - 1 -\epsilon) C(\lambda_i) + \beta + (1+\epsilon) C(\lambda) 
\leq
\alpha C(\lambda) + \beta
\end{align*}
and, hence, $\lambda \mapsto \tilde{\vec{x}} (\lambda)$ is an $(\alpha, \beta)$-approximate minimum-cost flow function.
\end{proof}

Not all the inequalities from Lemma~\ref{lem:mcfi} that bound the step size for the minimum-cost flow interpolation are applicable in practice. In particular, for the bound in \emph{(i)} we need the derivative of the objective function at the next breakpoint. By Theorem~\ref{thm:derivative_objective}, we can express this term as $\frac{d}{d \lambda} C (\lambda_i + \delta_i) = \vec{b}^\top \vec{\pi} (\lambda_i + \delta_i)$. Thus, this quantity depends on the optimal potentials at the next breakpoint and therefore we effectively need the optimal solution for the next breakpoint in order to compute the step size. Since this is not feasible, we want to find a rough estimate for this quantity that can be computed quickly in order to find a good estimate for the step size to the next breakpoint in reasonable time.
Similarly, the maximum and the Laplacian matrix $\vec{L}_{\lambda}^*$ in the condition~\emph{(iii)} are possibly hard to compute. In the following theorem, we will give conditions such that we can find an estimate for this term.

In order to find an estimate for the potentials $\vec{\pi} (\lambda_i + \delta_i)$, we need to introduce some additional notation. Recall the definition of the shortest path potential in \S~\ref{sec:functional_dependence}. The shortest path potential $\pi_v (\lambda)$ is defined as the length of a shortest path in the graph $\hat{G}$ that contains an additional backward edge for every edge with positive flow $x^*_e (\lambda)$ with edge weights $\nu_e = \mcost (x^*_e (\lambda))$ for forward edges and $\nu_e (\lambda) = - \mcost (x^*_e (\lambda))$ for backward edges. Now consider the graph $\tilde{G} = (V, \tilde{E})$, where $\tilde{E}$ contains every edge $e \in E$ from the original graph as well as backward edges for all edges. Then, for some given $M \geq 0$, we define the edge weights $\nu^M_e := \max \{ \mcost (M), - \mcost (-M)\}$ for all edges (forward and backward edges) and denote by $\nu_{v,w}^M$ the length of a shortest path in $\tilde{G}$ with respect to these weights.
For any two vertices $v, w$ denote by $P_{v,w}$ the shortest path with respect to the edge weights $\nu_e (\lambda)$ in $\hat{G}$ and by $Q_{v,w}$ the shortest path with respect to the edge weights $\nu_e^M$ in $\tilde{G}$.
If $M \geq |x^*_e (\lambda)|$ for all edges, then $\nu_e^M$ is an upper bound for $|\nu_e| = |\mcost (x^*_e (\lambda))|$. We therefore obtain
\begin{equation} \label{eq:potential_estimate}
\pi_w (\lambda) - \pi_v (\lambda) = \sum_{e \in P_{v,w}} \nu_e(\lambda) \leq \sum_{e \in P_{v,w}} \nu^M_e \leq \sum_{e \in Q_{v,w}} \nu^M_e = \nu^M_{v,w}
\end{equation}
and, hence, $\nu^M_{v,w}$ can be used as an upper bound for the potential differences of the shortest path potential.

Finally, we introduce the notion of a source-sink-decomposition. For some demand vector $\vec{b}$, we say a family of $J \in \N$ tuples $\mathcal{S}(\vec{b}) = \big( (s_j, t_j, r_j) \big)_{j=1, \dotsc, J}$ is a \emph{source-sink-decomposition of $\vec{b}$}, if $s_j, t_j \in V$ are vertices and $r_j \geq 0$ is some non-negative rate for every $j \in J$ such that
\[
b_v = 
\sum_{\substack{j \in \{1, \dotsc, J\} : \\ t_j = v}} r_j
-
\sum_{\substack{j \in \{1, \dotsc, J\} : \\ s_j = v}} r_j,
\]
i.e., $\mathcal{S}(\vec{b})$ contains source and sink vertices associated with demand rates that induce the demand vector~$\vec{b}$.

\begin{theorem} \label{thm:mcfi}
Let $\alpha > 1$, $\beta \geq 0$, and $0 < \epsilon < \alpha - 1$. Let $0 = \lambda_1 < \lambda_2 < \dots < \lambda_K = \lambda^{\max}$ be the breakpoints of the function $\lambda \mapsto \tilde{\vec{x}} (\lambda)$ defined in~\eqref{eq:mcfi:function}. If for every interval $[\lambda_i, \lambda_{i+1}]$ one of the following conditions holds true, then $\lambda \mapsto \tilde{\vec{x}} (\lambda)$ is an $(\alpha, \beta)$-approximate minimum-cost flow function.
\begin{enumerate}[(i)]
\item
	The objective function $\lambda \mapsto C(\lambda)$ is non-decreasing on $[\lambda_i, \lambda_{i+1}]$ and the step sizes~$\delta_i$ satisfy
	\[
	\delta_i \sum_{(s_j, t_j, r_j) \in \mathcal{S} (\vec{b})} r_j \nu^M_{s_j, t_j}
	\leq
	\frac{\alpha-1-\epsilon}{1+\epsilon} C(\lambda_i) + \frac{\beta}{1+\epsilon}
	,
	\]
	for $M := \sum_{j = 1}^J r_j (\lambda_i + \delta_i)$ for some source-sink-decomposition $\mathcal{S} (\vec{b})$.
\item
	For every $e \in E$, the marginal cost function is differentiable with $\mcost'(x) > 0$ for all $x \in \R$, convex for $x > 0$, and concave for $x < 0$. Further, the support does not change between $\lambda_i$ and $\lambda_{i+1}$, i.e., $S (\lambda) = S(\lambda_{i+1}$ for all $\lambda \in [\lambda_i, \lambda_{i+1}]$, and the step sizes $\delta_i$ satisfy
	\[
	\delta_i \sqrt{\vec{b}^{\top} \vec{L}^*_{B} \vec{b}}
	\leq
	2 \sqrt{2} \sqrt{\frac{\alpha - 1 - \epsilon}{1 + \epsilon} C(\lambda_i) + \frac{\beta}{1+\epsilon}}
	,
	\]
	where $\vec{L}_B$ is the Laplacian matrix with respect to the edge weights $c^B_e := \frac{1}{\max \{ \mcost' (B), \mcost'(-B)\}}$ if $e \in S(\lambda_i)$ and $c^B_e := 0$ if $e \notin S(\lambda_i)$, and $B := \max_{\lambda \in [\lambda_i, \lambda_{i+1}]} \frac{1}{2} \sum_{v \in V} |\lambda b_v + b_v^0|$ is the maximum value of the total inflow into the network in the interval $[\lambda_i, \lambda_{i+1}]$.
\end{enumerate}
\end{theorem}
\begin{proof}
Since $M = \sum_{j = 1}^J r_j (\lambda_i + \delta_i)$ is the total amount of flow entering the network for parameter $\lambda_{i+1} = \lambda_i + \delta_i$ and the edge cost are assumed to be non-decreasing, the value~$M$ is a trivial upper bound on the minimum-cost flow for parameter~$\lambda_{i+1}$ on any edge. Using the definition of a source-sink-decomposition and \eqref{eq:potential_estimate}, we obtain
\begin{align*}
\frac{d}{d \lambda} C (\lambda_i + \delta_i) 
&= \sum_{v \in V} b_v \pi_v (\lambda_i + \delta_i) \\
&= \sum_{j \in J} r_j \big( \pi_{t_j} (\lambda_i + \delta_i) - \pi_{s_j} (\lambda_i + \delta_i) \big) 
\stackrel{\eqref{eq:potential_estimate}}{\leq}
\sum_{j \in J} r_j 
\nu^M_{s_j, t_j}
\end{align*}
and, therefore, Lemma~\ref{lem:mcfi} implies that condition~\emph{(i)} is indeed sufficient.

Consider condition~\emph{(ii)}. Since the edge cost are non-decreasing, the total inflow into the network is a bound of the flow on every edge. In particular, $|x^*(\lambda)| \leq B = \max_{\lambda \in [\lambda_i, \lambda_{i+1}]} \frac{1}{2} \sum_{v \in V} |\lambda b_v + b_v^0|$. Using the convexity and concavity assumption, we get that $\mcost'(x^*_e(\lambda)) \leq \max \{ \mcost' (B), \mcost'(-B) \}$ for every edge $e \in E$. Thus, for every edge $e \in S(\lambda_i)$ in the support we get that  $c_e (\lambda) = \frac{1}{\mcost' (x^*(\lambda)} \geq \frac{1}{\max \{ \mcost' (B), \mcost'(-B) \}} = C_B$ and for every edge $e \notin S(\lambda_i)$ we have $c_e(\lambda) = 0 = c_e^B$. With Rayleigh's monotonicity law from Theorem~\ref{thm:thomson_rayleigh} we obtain
\[
\vec{b}^{\top} \vec{L}^*_{\lambda} \vec{b} \leq \vec{b}^{\top} \vec{L}^*_{B} \vec{b}
\]
for all $\lambda \in [\lambda_i, \lambda_{i+1}]$. Hence, again with Lemma~\ref{lem:mcfi}, the claim follows.
\end{proof}

\section{Computational Study} \label{sec:computational_study}

In this section, we study the applicability of the methods from \S~\ref{sec:generalcost} for the approximate parametric computation of minimum-cost flows to real-world instances in practice.
In particular, we use a Python implementation of the algorithm in order to solve parametric problems on real-world traffic and gas instances. 
The traffic networks are directed networks with polynomial edge costs. We compute traffic equilibria in these networks which are equivalent to minimum-cost flows by the Beckmann transformation~\cite{beckmann1956}. The gas instances consist of undirected networks and the gas flow equilibria in these networks can be reformulated as minimum-cost flows with polynomial cost functions.

We test both variants of our algorithm described in \S~\ref{sec:generalcost}, the marginal cost approximation~(MCA) and the minimum-cost flow interpolation~(MCFI). In the former variant, we interpolate the marginal cost functions $\mcost$ with linear splines using the step sizes from Theorem~\ref{thm:mca} and solve the resulting instances with the parametric algorithm for instances with piecewise quadratic cost functions from~\S~\ref{sec:quadratic} . The latter variant computes approximate solutions for given parameter values $\lambda_1, \lambda_2, \dots$ with step sizes defined by Theorem~\ref{thm:mcfi} and interpolate the returned minimum-cost flow functions.

\subsection{Implementation}

We use a Python~3 implementation of the described algorithms. The networks are implemented based on edge and vertex lists. For shortest path computations we use the implementation of Dijkstra's algorithm from the SciPy package~\cite{scipy}. For most linear algebra computations, we use sparse arrays from the Scipy package. In particular, we use an implementation from the same package of the Cholesky decomposition to obtain, maintain, and update the inverse of the involved Laplacian matrices.

For the minimum-cost flow interpolation we implement a variant of the Frank-Wolfe algorithm~\cite{frank1956algorithm}, also known as the \emph{conditional gradient method}. We use the parallel tangents (PARTAN) method (see Florian et al.~\cite{florian1987efficient}) to improve the convergence rate of the algorithm. The Frank-Wolfe algorithm requires solutions of subproblems which are in our case either classical minimum-cost flow problems with linear cost or, in the case of directed, single-commodity networks, shortest paths computations. For both types of subproblems we use the SciPy package; the Simplex algorithm to solve the former and Dijkstra's algorithm to solve the latter. 

The implementation together with a documentation is publically available online as a Python package name \emph{paminco} and can be found on the paminco GitHub page~\cite{paminco-github}.

\subsection{Instances}

\subsubsection{Traffic networks from the Transportation Networks Library}

The first set of instances we consider are traffic networks based on real road networks of cities or larger areas provided by the \emph{Transportation Networks} library~\cite{transportation}. The networks are given in the TNTP-file format that we convert into a minimum-cost flow instance. The TNTP-files provide the network structure containing vertices, edges, and edge cost functions. The edge cost functions model the travel time required to traverse an edge depending on the congestion created by the flow on the edge and are given in the form
\[
c_e (x) = \mathrm{fft}_e \Bigg( 1 + B_e \cdot \bigg( \frac{x}{\mathrm{cap}_e} \bigg)^4 \Bigg),
\]
where the coefficients model the free flow time $\mathrm{fft}_e$ on the edge (that is the time it takes to traverse the edge without any congestion), the capacity $\mathrm{cap}_e$ of the edge, and a coefficient $B_e$ that models the effects of congestion on this edge. The TNTP-file format also specifies demands of a multi-commodity setting with several commodities spread out through the network. In this computational study, we will only consider the single-commodity case. Therefore, we replace the commodities by a single commodity with random source $s$ and sink vertex $t$ and a demand rate $r$ that is a fraction of the overall demand $D$ of all commodities specified in the TNTP-files. Concretely, we use $r = \frac{D}{2}$ for the Berlin networks and $r = \frac{D}{10}$ for the American networks in order to achieve a sufficient congestion on the edges. Then, we compute parametric traffic flows for demand rates scaled by the parameter $\lambda \in [0,1]$. We repeat the computation for several random source and sink pairs.
In this setting, we want to compute Wardrop equilibria with respect to these edge costs. By the Beckmann transformation~\cite{beckmann1956}, this is equivalent to solving the minimum-cost flow problem
\begin{align*}
\min \sum_{e \in E} \int_0^{x_e} c_e (s) ds &= 
\sum_{e \in E} \mathrm{fft}_e \Bigg( x_e + \frac{B_e}{5} \cdot \frac{x_e^5}{\mathrm{cap}_e^4} \Bigg) \\
\text{s.t.} \qquad
\vec{\Gamma} \vec{x} &= \lambda \vec{b} \\
\vec{x} &\geq \vec{0},
\end{align*}
where $\vec{b} = (b_v)_{v \in V}$ with $b_s = -1$, $b_t = 1$, and $b_v = 0$ for all $v \neq s, t$.

We also use some preprocessing routines on the network that remove isolated and artificial vertices as well as some edges with zero cost functions in order as our algorithms require a connected graph and strictly convex cost functions. The vertices and edges removed in these preprocessing steps are only used as artificial entries and exits of the network and their removal does not affect the general structure of the networks. The network sizes given in Table~\ref{tab:studyresults} contain the number of vertices and edges of the networks after this preprocessing step.

\subsubsection{Gas networks from the GasLib}

The second set of test instances we use are gas networks from the GasLib~\cite{SABHJKKIOSSS17}. This library contains several schematic and real-world gas networks in different sizes. Gas networks are undirected networks (and can be therefore treated like electrical networks) and the edges model pipes, compressors, valves, and other components of a gas pipeline network. Further, the library provides information of gas inflow and outflow rates in several scenarios.

Gas flows through a pipe $e = (v, w)$ can be modeled by the \emph{Weymouth equations}~\cite{weymouth1912problems}
\[
p_w^2 - p_v^2 = \beta_e x_e |x_e|,
\]
where $p_v^2$ is the pressure value at vertex~$v$ and $\beta_e > 0$ is an edge-dependent coefficient modeling the resistance of the pipe~$e$. With the substitution $\pi_v = p_v^2$ we see that a gas flow in a network consisting only of pipes 
can be interpreted as a solution to an undirected minimum-cost flow problem 
with marginal cost functions $\mcost (x) = \beta_e x |x|$.
We neglect all components of the network other  than pipes (such as compressors and valves)  by contracting the respective edges. This can be interpreted as modeling the network where all compressors are shut down and all valves are open. The coefficients $\beta_e$ modeling the resistance of the edges can be obtained from physical properties of the pipes, such as length, roughness, and diameter. These properties are also provided in the GasLib data. For more details and the exact formula for the computation of the coefficients $\beta_e$, see, e.g., Pfetsch et al.~\cite{pfetsch2015validation}.

For every gas network, we create $N=20$ parametric instances as follows. First, we choose one fixed demand scenario from the data corresponding to some base demand $\bar{\vec{b}}$. Then, we choose $N=20$ pairs consisting of one source vertex $s$ and one sink vertex $t$. With these two vertices, we define an affine-linear demand function with an offset demand $\bar{\vec{b}}$ and a demand direction that corresponds to $s$-$t$-flows of rate $r := \frac{1}{2} \sum_{v \in V} |\bar{b}_v|$. Formally, define the demand vector $\vec{b}$ with $b_s = - r$, $b_t = r$, and $b_v = 0$ for all $v \neq s, t$. Then we consider the demand function
$
\lambda \mapsto \lambda \vec{b} + \bar{\vec{b}}
$
and compute the gas flows for all parameters $\lambda \in [0, 1]$. Hence, for every network we consider $N=20$ instances with the parametric setting where we start with the basic demand and then increase the demand for one (random) $s$-$t$-pair. 
Since the parametric computation does not start with the zero-demand, we need to compute an initial solution for the MCA method. In the computational experiments, the MCA algorithm itself for obtaining this initial solution. Before solving the actual instance, we first compute the solution to the problem with the linear demand function $\lambda \mapsto \lambda \bar{\vec{b}}$. In this initial run, that we also refer to as Phase~I, all cost functions are homogeneous and the demand function is linear. Thus, the initial solution for this problem is trivial as it is the zero flow associated with the zero potential. After solving the initial problem for $\lambda \in [0,1]$, we can use its solution for $\lambda = 1$ as the initial solution for the actual run of the MCA algorithm, that we also refer to as Phase~II.

For the minimum-cost flow interpolation approach, we need to use a step size rule from Theorem~\ref{thm:mcfi}. The first rule of this corollary requires the objective function $\lambda \mapsto C(\lambda)$ to be non-decreasing, which is not guaranteed in the case of affine linear demands.
Therefore, we want to use the second rule from Theorem~\ref{thm:mcfi}\emph{(ii)}. Since we consider an undirected problem, the supports are always the same for all parameter values. However, the marginal cost functions $\mcost (x) = \beta_e x |x|$ do not have strictly positive derivative needed to apply this step size rule. We solve this problem by adding a regularization term of $\zeta x$ to the marginal cost. This guarantees that the marginal cost have a positive derivative. The following lemma proves, that for $\zeta$ small enough, we can compute $(\alpha, \beta)$-approximate solutions to any precision.

\begin{lemma} \label{lem:gasapx}
Let $\lambda \mapsto \vec{x}^*(\lambda)$ be the minimum-cost flow function in an instance with marginal cost functions $\mcost (x) = \beta_e x |x|$ and denote by $C(\lambda)$ the cost of the flow~$\vec{x}^*(\lambda)$. Denote by $\beta^* = \min_{e \in E} \beta_e$. Further, let $\lambda \mapsto \tilde{\vec{x}}^* (\lambda)$ be the minimum-cost flow for the same instance but with marginal cost functions $\tilde{\mcost} (x) := \beta_e x |x| + \zeta x$ and denote the cost of this flow with respect to the orginial cost functions by $\tilde{C} (\lambda)$. Let $\alpha > 1$, $\beta > 0$, $\lambda^{\max} > 0$, and $x^{\max} := \max_{\lambda \in [0, \lambda^{\max}]} \sum_{v \in V} \frac{|b_v(\lambda)|}{2}$. If $\zeta \leq 2 \sqrt{\frac{(\alpha - 1) \beta \beta^*}{m x^{\max}}}$, then
\begin{equation} \label{eq:lem:gasapx}
\tilde{C}(\lambda) \leq \alpha C(\lambda) + \beta 
\quad \text{for all } \lambda \in [0, \lambda^{\max}],
\end{equation}
i.e., $\tilde{\vec{x}}^* (\lambda)$ is an $(\alpha, \beta)$-approximate minimum-cost flow.
\end{lemma}
\begin{proof}
By assumption, we obtain
\begin{align*}
| \mcost&(x)- \tilde{\mcost}(x) | 
= \zeta |x| 
\leq
2 \sqrt{\frac{(\alpha - 1) \beta \beta^*}{m x^{\max}} } |x|
\\
&= (\alpha -1 ) \beta^* x^2 + \frac{\beta}{m x^{\max}} 
-  \Bigg( \!\! (\alpha-1) \beta^* x^2 - 2 \sqrt{\frac{(\alpha - 1) \beta \beta^*}{m x^{\max}} } |x| +  \frac{\beta}{m x^{\max}} \Bigg) \\
&=  (\alpha -1 ) \frac{\beta^*}{\beta_e} |\mcost (x)| + \frac{\beta}{m x^{\max}} - \Bigg( \!\! \sqrt{(\alpha - 1) \beta^*} |x|- \sqrt{\frac{\beta}{m x^{\max}} } \Bigg)^2 
\\
&\leq (\alpha -1 ) |\mcost (x)| + \frac{\beta}{m x^{\max}}
\end{align*}
and, thus, with Lemma~\ref{lem:mca} the claim follows.
\end{proof}

We use the original instance for the results of the marginal cost approximation and the instance with cost functions marginal cost functions $\tilde{\mcost} (x) := \beta_e x |x| + \zeta x$ for the minimum-cost flow interpolation. For the latter, we choose a value of $\zeta$ such that \eqref{eq:lem:gasapx} holds for $\alpha = 1 + 10^{-5}$ and $\beta = 10^{-5}$. We then adjust the values for $\alpha$ and $\beta$ in the computation of the step sizes from Theorem~\ref{thm:mcfi}\emph{(ii)} for the interpolation such that we can guarantee the overall approximation for $\alpha = 1.01$ and $\beta = 1$.

\subsection{Computational Experiments and Results}

We apply both algorithmic approaches from \S~\ref{sec:generalcost}, namely marginal cost approximation (MCA) and minimum-cost flow interpolation (MCFI), to five traffic networks from the Transporation Networks library and six different gas networks from the GasLib library. Further, we compute fixed solutions for the parameter $\lambda = 1$ with the Frank-Wolfe algorithm (FW) to compare the computation times of the parametric algorithms to the effort of computing a solution for a fixed demand. All computational experiments are carried out on a Google Cloud virtual machine of type \emph{c2-standard-8}. The virtual machine provides eight virtual CPUs with up to 3.8~GHz CPU computing frequency each and 32~GB~RAM.
The Python implementation of both algorithms can be found on GitHub~\cite{paminco-github}. 
Both algorithms compute $(\alpha, \beta)$-approximate minimum-cost flows. For all computational experiments we fix $\alpha = 1.01$ and $\beta = 1$. 

\subsubsection{$\fwprec$-trade-off for the minimum-cost flow interpolations}

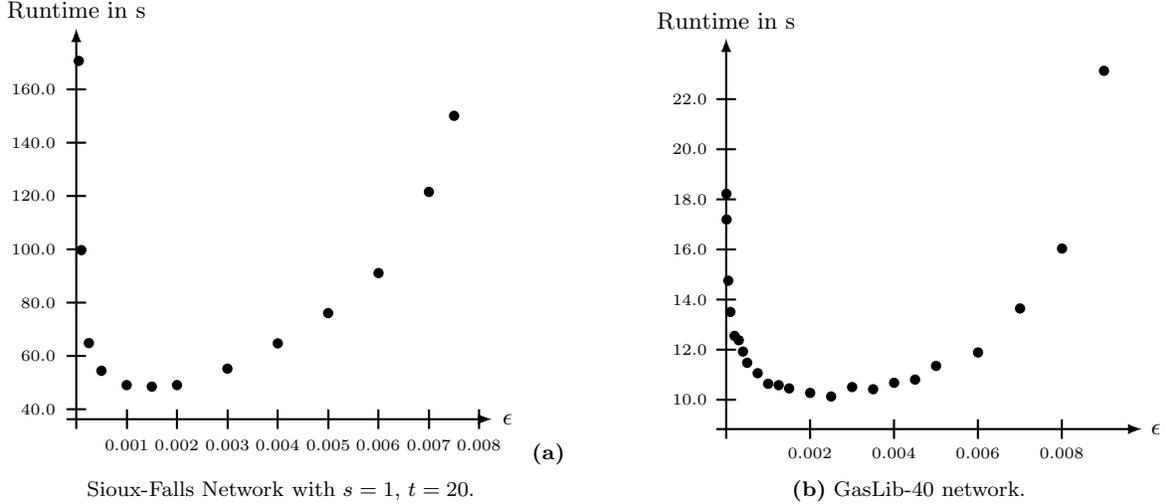
\begin{figure}
    \centering
    \begin{minipage}[b]{.47\linewidth}
    \begin{center}
    \begin{tikzpicture}[scale=1.3]
    \draw[thick, ->] (-.1, 0) -- (4.250000, 0) node[right] {\footnotesize $\epsilon$};
    \draw[thick, ->] (0, -.1) -- (0, 4.000000) node[above] {\footnotesize Runtime in s};
    
    \begin{scope}[xscale=515.151515]
    \foreach \x in {0.001, 0.002, 0.003, 0.004, 0.005, 0.006, 0.007, 0.008} {
        \draw[thick] (\x, .1) -- (\x, -.1) node[below] {\tiny \x};
    }
    \end{scope}
    
    \begin{scope}[yscale=0.027281]
    \foreach \y in {40.0, 60.0, 80.0, 100.0, 120.0, 140.0, 160.0} {
        \draw[thick] (.1, \y-36.248431) -- (-.1, \y-36.248431) node[left] {\tiny \y};
    }
    \end{scope}
    
    \foreach \point in {(0.025758, 3.666667), (0.051515, 1.730614), (0.128788, 0.779001), (0.257576, 0.495745), (0.515152, 0.349130), (0.772727, 0.333333), (1.030303, 0.349455), (1.545455, 0.517360), (2.060606, 0.776270), (2.575758, 1.086183), (3.090909, 1.495787), (3.606061, 2.326631), (3.863636, 3.104994)} {
        \fill \point circle(1.5pt);
    }
    \end{tikzpicture}
    {\scriptsize \textbf{(a)} Sioux-Falls Network with $s=1$, $t=20$.}
    \end{center}
    \end{minipage}
    \hspace{0.02\linewidth}
    \begin{minipage}[b]{.47\linewidth}
    \begin{center}
    \footnotesize
    \begin{tikzpicture}[scale=1.3]
    \draw[thick, ->] (-.1, 0) -- (4.250000, 0) node[right] {\footnotesize $\epsilon$};
    \draw[thick, ->] (0, -.1) -- (0, 4.000000) node[above] {\footnotesize Runtime in s};
    
    \begin{scope}[xscale=429.292929]
    \foreach \x in {0.002, 0.004, 0.006, 0.008} {
        \draw[thick] (\x, .1) -- (\x, -.1) node[below] {\tiny \x};
    }
    \end{scope}
    
    \begin{scope}[yscale=0.256371]
    \foreach \y in {10.0, 12.0, 14.0, 16.0, 18.0, 20.0, 22.0} {
        \draw[thick] (.1, \y-8.829377) -- (-.1, \y-8.829377) node[left] {\tiny \y};
    }
    \end{scope}
    
    \foreach \point in {(0.002146, 2.407377), (0.004293, 2.144549), (0.021465, 1.518817), (0.042929, 1.198900), (0.085859, 0.953739), (0.128788, 0.909161), (0.171717, 0.793097), (0.214646, 0.678595), (0.321970, 0.572028), (0.429293, 0.463533), (0.536616, 0.449296), (0.643939, 0.416453), (0.858586, 0.370328), (1.073232, 0.333333), (1.287879, 0.430296), (1.502525, 0.407565), (1.717172, 0.473856), (1.931818, 0.506511), (2.146465, 0.646758), (2.575758, 0.784038), (3.005051, 1.234326), (3.434343, 1.847880), (3.863636, 3.666667)} {
        \fill \point circle(1.5pt);
    }
    \end{tikzpicture}
    
    {\scriptsize \textbf{(b)} GasLib-40 network.}
    \end{center}
    \end{minipage}
    \caption{Runtimes of the minimum-cost flow interpolation algorithm for different values of $\fwprec$.}
    \label{fig:epsilontradeoff}
\end{figure}

The minimum-cost flow interpolation has an additional parameter $0 < \fwprec < \alpha - 1$ that governs the trade-off between the approximation quality of the Frank-Wolfe algorithm for the fixed parameter values and the approximation quality of the interpolation of the minimum-cost flows. 
We begin the computational study by trying to find an optimal value for $\fwprec$, both for the directed traffic and the undirected gas networks. To this end, we use the Sioux-Falls network from the Transportation Networks library and the GasLib-40 network from the GasLib with fixed demands and compute the parametric solution with the minimum-cost flow interpolation for varying values of $\fwprec$. 

Figure~\ref{fig:epsilontradeoff} shows results of the trade-off. We see that for low and high values of $\fwprec$ the runtime of the parametric algorithm increases. For low values of $\fwprec$ the runtime increases as the convergence threshold for the Frank-Wolfe algorithm for the fixed demands decreases and therefore the runtime of the Frank-Wolfe algorithm increases. For high values of $\fwprec$ the runtime of the Frank-Wolfe algorithm is low, however, as we can see from the step size formulas from Lemma~\ref{lem:mcfi}, the step sizes decrease and, hence, the number of breakpoints increases resulting in more calls of the Frank-Wolfe algorithm in total.
The trade-off yields an optimal value $\fwprec = 0.0015$ for the traffic network and an optimal value of $\fwprec = 0.0025$ for the gas network. We will use these values for all traffic and gas networks, respectively, in all subsequent computations, although it should be noted that the optimal value also depends on the network and choice of demands. However, it is not feasible to perform this trade-off for every network prior to the actual computation since it already requires solving the parametric problem multiple times rendering the purpose of the whole trade-off pointless. Hence, we use these two values as heuristic value for $\fwprec$ for all computations.

\subsubsection{Parametric computation of Wardrop equilibria in the Sioux-Falls network}

\begin{figure}[t]
\centering
\begin{minipage}[b]{.3\linewidth}
\footnotesize
\centering

\begin{tikzpicture}[scale=.85]
\begin{scope}[scale=.15]
\foreach \vertex/\x/\y in {
	 1 /  5 / 51, 
	 2 / 32 / 51,
	 3 /  5 / 44,
	 4 / 13 / 44,
	 5 / 22 / 44,
	 6 / 32 / 44,
	 7 / 42 / 38,
	 8 / 32 / 38,
	 9 / 22 / 38,
	10 / 22 / 32,
	11 / 13 / 32,
	12 /  5 / 32,
	13 /  5 /  5,
	14 / 13 / 19,
	15 / 22 / 19,
	16 / 32 / 32,
	17 / 32 / 26,
	18 / 42 / 32,
	19 / 32 / 19,
	20 / 32 /  5,
	21 / 22 /  5,
	22 / 22 / 13,
	23 / 13 / 13,
	24 / 13 /  5 
} {
	\node (\vertex) at (\x, \y) {\vertex};
}

\tikzset{auto rotate/.style={auto=right,->,
to path={let \p1=(\tikztostart),\p2=(\tikztotarget),
\n1={atan2(\y2-\y1,\x2-\x1)},\n2={\n1-20},\n3={\n1+200}
in (\tikztostart.\n2) -- (\tikztotarget.\n3) \tikztonodes}}}

\draw[->, auto rotate]
( 1 ) edge ( 2 )
( 1 ) edge ( 3 )
( 2 ) edge ( 1 )
( 2 ) edge ( 6 )
( 3 ) edge ( 1 )
( 3 ) edge ( 4 )
( 3 ) edge ( 12 )
( 4 ) edge ( 3 )
( 4 ) edge ( 5 )
( 4 ) edge ( 11 )
( 5 ) edge ( 4 )
( 5 ) edge ( 6 )
( 5 ) edge ( 9 )
( 6 ) edge ( 2 )
( 6 ) edge ( 5 )
( 6 ) edge ( 8 )
( 7 ) edge ( 8 )
( 7 ) edge ( 18 )
( 8 ) edge ( 6 )
( 8 ) edge ( 7 )
( 8 ) edge ( 9 )
( 8 ) edge ( 16 )
( 9 ) edge ( 5 )
( 9 ) edge ( 8 )
( 9 ) edge ( 10 )
( 10 ) edge ( 9 )
( 10 ) edge ( 11 )
( 10 ) edge ( 15 )
( 10 ) edge ( 16 )
( 10 ) edge ( 17 )
( 11 ) edge ( 4 )
( 11 ) edge ( 10 )
( 11 ) edge ( 12 )
( 11 ) edge ( 14 )
( 12 ) edge ( 3 )
( 12 ) edge ( 11 )
( 12 ) edge ( 13 )
( 13 ) edge ( 12 )
( 13 ) edge ( 24 )
( 14 ) edge ( 11 )
( 14 ) edge ( 15 )
( 14 ) edge ( 23 )
( 15 ) edge ( 10 )
( 15 ) edge ( 14 )
( 15 ) edge ( 19 )
( 15 ) edge ( 22 )
( 16 ) edge ( 8 )
( 16 ) edge ( 10 )
( 16 ) edge ( 17 )
( 16 ) edge ( 18 )
( 17 ) edge ( 10 )
( 17 ) edge ( 16 )
( 17 ) edge ( 19 )
( 18 ) edge ( 7 )
( 18 ) edge ( 16 )
( 18 ) edge ( 20 )
( 19 ) edge ( 15 )
( 19 ) edge ( 17 )
( 19 ) edge ( 20 )
( 20 ) edge ( 18 )
( 20 ) edge ( 19 )
( 20 ) edge ( 21 )
( 20 ) edge ( 22 )
( 21 ) edge ( 20 )
( 21 ) edge ( 22 )
( 21 ) edge ( 24 )
( 22 ) edge ( 15 )
( 22 ) edge ( 20 )
( 22 ) edge ( 21 )
( 22 ) edge ( 23 )
( 23 ) edge ( 14 )
( 23 ) edge ( 22 )
( 23 ) edge ( 24 )
( 24 ) edge ( 13 )
( 24 ) edge ( 21 )
( 24 ) edge ( 23 );
\end{scope}
\end{tikzpicture}
{\scriptsize \textbf{(a)} Sioux-Falls network.}
\end{minipage}
\hspace{0.03\linewidth}
\begin{minipage}[b]{.60\linewidth}
\footnotesize
\centering
\begin{tikzpicture}[scale=.95]
\draw[thick, ->] (-.5,0) -- (6.5,0) node[right] {Inst.};
\draw[thick, ->] (0,-.5) -- (0,6) node[above] {Runtime in s};
\foreach \y in {10,20,30,40,50,60,70,80,90,100} { 
	\draw[thick] (.1, \y / 110 * 6.000 ) -- (-.1, \y / 110 * 6.000 ) node[left] {$\y$}; 
}

\draw[thick, dashed, black]
	(0,0.290 / 110 * 6.000) -- (6.25,0.290 / 110 * 6.000);
\draw[thick, dashdotted, black]
	(0,48.459 / 110 * 6.000) -- (6.25,48.459 / 110 * 6.000);

\foreach \point in {(0.120, 0.025), (0.240, 0.014), (0.360, 0.017), (0.480, 0.017), (0.600, 0.016), (0.720, 0.016), (0.840, 0.013), (0.960, 0.016), (1.080, 0.017), (1.200, 0.015), (1.320, 0.015), (1.440, 0.019), (1.560, 0.014), (1.680, 0.017), (1.800, 0.015), (1.920, 0.015), (2.040, 0.011), (2.160, 0.017), (2.280, 0.015), (2.400, 0.012), (2.520, 0.019), (2.640, 0.016), (2.760, 0.018), (2.880, 0.019), (3.000, 0.015), (3.120, 0.017), (3.240, 0.015), (3.360, 0.015), (3.480, 0.018), (3.600, 0.018), (3.720, 0.013), (3.840, 0.016), (3.960, 0.014), (4.080, 0.018), (4.200, 0.017), (4.320, 0.012), (4.440, 0.015), (4.560, 0.017), (4.680, 0.018), (4.800, 0.014), (4.920, 0.016), (5.040, 0.015), (5.160, 0.013), (5.280, 0.014), (5.400, 0.015), (5.520, 0.016), (5.640, 0.013), (5.760, 0.015), (5.880, 0.018), (6.000, 0.015)} {
	\fill[black] \point circle(2pt);
}
\foreach \point in {(0.120, 3.826), (0.240, 0.733), (0.360, 2.366), (0.480, 3.766), (0.600, 3.468), (0.720, 3.136), (0.840, 0.626), (0.960, 2.508), (1.080, 3.782), (1.200, 0.926), (1.320, 1.152), (1.440, 2.033), (1.560, 0.928), (1.680, 3.362), (1.800, 3.477), (1.920, 2.901), (2.040, 0.465), (2.160, 3.048), (2.280, 2.897), (2.400, 0.537), (2.520, 3.281), (2.640, 1.881), (2.760, 3.212), (2.880, 2.844), (3.000, 3.935), (3.120, 3.046), (3.240, 4.514), (3.360, 4.124), (3.480, 5.699), (3.600, 2.011), (3.720, 0.585), (3.840, 3.778), (3.960, 0.887), (4.080, 5.498), (4.200, 3.462), (4.320, 0.452), (4.440, 1.016), (4.560, 3.361), (4.680, 2.173), (4.800, 1.292), (4.920, 4.600), (5.040, 1.116), (5.160, 0.983), (5.280, 1.336), (5.400, 3.750), (5.520, 3.773), (5.640, 0.582), (5.760, 4.448), (5.880, 5.599), (6.000, 2.985)} {
	\fill[white, draw=black] \point circle(2pt);
}
\begin{scope}[shift={(-.55,-1.25)}]
\draw (-.45,.5) rectangle (8.1, -.5);
\fill[black] (0,.25) circle(2pt) node[right=.25cm, black] {MCA runtime};
\draw[thick, dashed, black]    (-.25,-.25) -- (.25, -.25) node[right, black] (xyz) {avg. MCA time {\tiny($0.3\,$s)}};
\fill[white, draw=black] (4.05,.25) circle(2pt) node[right=.25cm, black] {MCFI runtime};
\draw[thick, dashdotted, black]
(3.8,-.25) -- (4.3, -.25) node[right, black] {avg. MCFI time {\tiny($48.5\,$s)}};
\end{scope}
\end{tikzpicture}
\\
{\scriptsize \textbf{(b)} Runtimes of both algorithms for random instances.}
\end{minipage}
\caption{Results of the computational study for the Sioux-Falls traffic network.}
\label{fig:siouxfalls}
\end{figure}
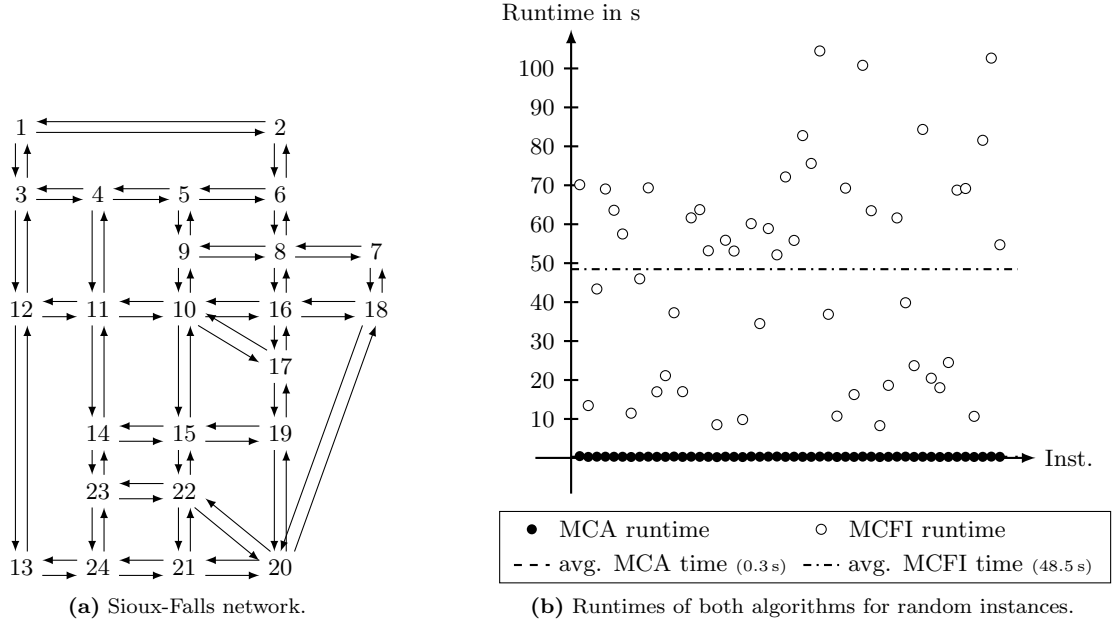%
\begin{figure}[t]
\centering
\begin{minipage}[b]{.95\linewidth}
\footnotesize
\centering
\input{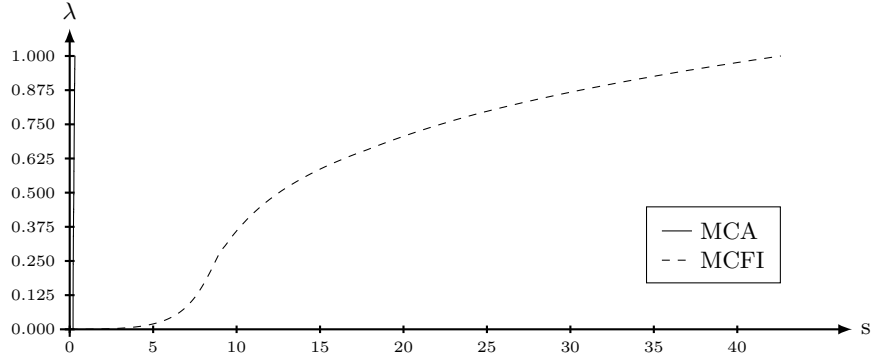}
\\
{\scriptsize \textbf{(a)} The progress over time for both algorithms for the instance with $s=1$ and $t=24$.}
\end{minipage}
\begin{minipage}[b]{.95\linewidth}
\footnotesize
\centering
\input{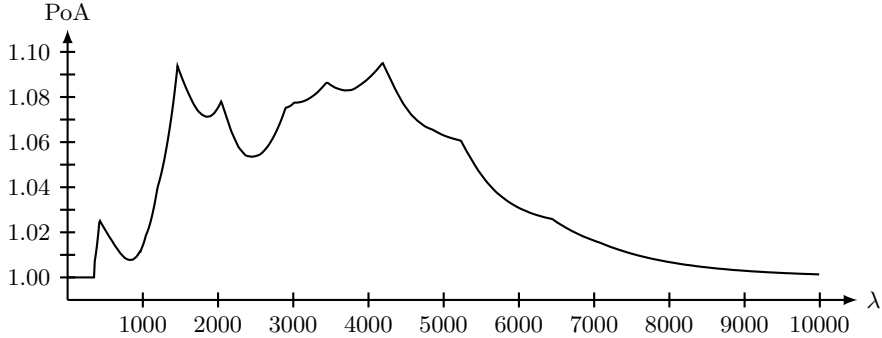}
\\
{\scriptsize \textbf{(b)} The price of anarchy curve for a demand rate of $r=10$ routed between the vertices $s=20$ and $t=2$ for $\lambda \in [0,10000]$, replicating a result of O'Hare et~al.~\cite{ohare2016}.}
\end{minipage}
\caption{Results for two fixed instances of the Sioux Falls network. (a) The progress over time of MCA and MCFI for the fixed instance with $s=204$ and $t = 248$. (b) Replication of the price of anarchy curve for a fixed instance as done by O'Hare et~al.~\cite{ohare2016}.}
\label{fig:siouxfalls2}
\end{figure}%

As a specific example for the traffic networks, we consider the Sioux-Falls network from Figure~\ref{fig:siouxfalls}(a). In $k=50$ iterations, we draw a random source and sink vertex $s_k, t_k$ from all vertices and use a demand rate of $r = \frac{D}{10}$, where $D$ is the total demand of the original data. This leads to a demand rate of $r = 36060$ in this concrete setting which ensures that the network is sufficiently congested. We use the marginal cost approximation (MCA) and the minimum-cost flow interpolation (MCFI) to obtain a parametric solution for $\lambda \in [0,1]$ of all $50$~instances. The results are shown in Figure~\ref{fig:siouxfalls}(b). The MCA algorithm has an average runtime of $290$ms over fifty random instances with the runtimes (black dots in Figure~\ref{fig:siouxfalls}(b)) ranging from $200$ms to $460$ms. The MCFI algorithm has an average runtime of $48.5$s over fifty random instances with the runtimes (white dots in Figure~\ref{fig:siouxfalls}(b)) ranging from $8.3$s to $104.5$s. For all random $s$-$t$-pairs the MCA algorithm has a huge advantage over the MCFI algorithm; the MCFI algorithm has on average about $160$-times longer runtimes on the random instances.
Figure~\ref{fig:siouxfalls2}(a) shows the progress of both algorithms over time for one particular instance (with source $s=1$ and $t=24$): The graph shows how far the parametric output functions have been computed (in terms of the parameter $\lambda$) after a given runtime in seconds. We see that the MCA algorithm makes steady and fast progress, while the MCFI needs a long time before it makes progress and the progress is decreasing for higher values of $\lambda$. The varying progress of the MCFI algorithm can be explained by the step size rule used for the algorithm: Since the support can change in the directed traffic networks, we use the step rule from Theorem~\ref{thm:mcfi}(i) that makes no assumptions on the support. For small values of $\lambda$ the value $C (\lambda_i)$ is very small resulting in small steps $\delta_i$. For high values of $\lambda$, the estimate of the derivative $\frac{d}{d \lambda} C (\lambda_i + \delta_i)$ developed in Theorem~\ref{thm:mcfi} is very large, leading again to smaller steps.

Finally, we use the MCA algorithm to replicate a result of O'Hare et~al.~\cite{ohare2016} who  studied the price of anarchy, i.e., the cost of the Wardrop equilibrium divided by the cost of a social optimal solution, for varying demands. Since both the equilibrium and the optimal flow can be computed as a minimum-cost flow, we can use the MCA algorithm to compute the parametrized equilibrium and optimum flows and, thus, the corresponding cost and the price of anarchy curve. O'Hare et~al. chose a setting where a demand rate of $r=10$ is routed between the vertices $s=20$ and $t=3$ and computed the price of anarchy curve for parameters $\lambda \in [0, 10000]$. Our implementation of the MCA algorithm can compute the equilibrium and optimum flows and also the respective cost, resulting in the curve shown in Figure~\ref{fig:siouxfalls2}(b). The equilibrium solution is computed in $13.3$s and the optimal flows in $20.0$s. Since the output functions of the MCA algorithms have breakpoints at exactly the points where the (marginal) cost functions of the network have breakpoints, we can in particular can identify the values of $\lambda$ where the flows on the edges becomes positive or zero. These points are called route transition points in the paper by O'Hare et~al. and can also be easily obtained from the output of the MCA algorithm. 
This example shows that the MCA algorithm can be used in practice to analyze price of anarchy curves with reasonable computing time (in our case approximately $30$ seconds for the Sioux-Falls network).

\subsubsection{Parametric computation of gas flows in the GasLib-40 network}

\begin{figure}[t]
\centering
\begin{minipage}[b]{.32\linewidth}
\footnotesize
\centering
\begin{tikzpicture}[scale=.8]
\begin{scope}[scale=.1]
\foreach \vertex/\x/\y in {%
 1 / 15.4 / 56.8,
 2 /  0.6 / 53.3,
 3 / 54.5 / 78.5,
 4 /  8.5 /  0.2,
 5 / 13.1 / 42.9,
 6 / 17.8 / 54.6,
 7 / 20.3 / 21.6,
 8 / 20.3 / 14.4,
 9 / 17.4 / 11.3,
10 / 18.4 / 11.5,
11 / 23.7 / 21.1,
12 / 10.2 / 18.4,
13 / 40.8 / 51.7,
14 / 44.4 / 48.9,
15 / 25.4 /  0.2,
16 / 26.3 / 44.3,
17 / 27.9 / 45.0,
18 / 16.0 / 44.1,
19 / 63.0 / 41.5,
20 / 21.7 / 19.4,
21 / 10.7 / 15.9,
22 / 47.2 / 67.4,
23 / 23.2 / 25.8,
24 / 24.4 /  3.1,
25 / 11.4 /  3.6,
26 / 18.2 / 57.7,
27 / 24.2 /  3.8,
28 / 11.5 / 22.3,
29 / 46.4 / 74.0,
30 /  7.4 / 54.7,
31 /  8.7 / 49.3,
32 / 46.3 / 67.1,
33 / 53.2 / 78.3,
34 / 21.3 / 42.4
} {
	\node[circle, fill,inner sep=1pt,outer sep=.5pt] (\vertex) at (\x, \y) {};
}

\tikzset{auto rotate/.style={auto=right,
to path={let \p1=(\tikztostart),\p2=(\tikztotarget),
\n1={atan2(\y2-\y1,\x2-\x1)},\n2={\n1-20},\n3={\n1+200}
in (\tikztostart.\n2) -- (\tikztotarget.\n3) \tikztonodes}}}

\draw[auto rotate]
(  1 ) edge (  6 )
( 14 ) edge ( 19 )
( 34 ) edge ( 16 )
( 16 ) edge ( 17 )
( 17 ) edge ( 13 )
( 34 ) edge ( 28 )
( 28 ) edge ( 12 )
( 12 ) edge ( 21 )
( 28 ) edge (  7 )
(  7 ) edge ( 23 )
( 21 ) edge (  9 )
( 34 ) edge (  6 )
(  9 ) edge ( 10 )
(  9 ) edge ( 25 )
( 10 ) edge ( 27 )
( 25 ) edge (  4 )
( 27 ) edge ( 24 )
( 24 ) edge ( 15 )
( 10 ) edge (  8 )
(  8 ) edge ( 20 )
( 20 ) edge (  7 )
( 20 ) edge ( 11 )
(  6 ) edge ( 26 )
( 11 ) edge ( 23 )
( 34 ) edge ( 23 )
( 34 ) edge ( 18 )
( 18 ) edge ( 31 )
( 31 ) edge ( 30 )
( 31 ) edge (  5 )
(  5 ) edge ( 18 )
( 31 ) edge (  2 )
(  3 ) edge ( 22 )
( 22 ) edge ( 32 )
(  3 ) edge ( 33 )
( 29 ) edge ( 33 )
( 29 ) edge ( 22 )
( 13 ) edge ( 14 )
( 13 ) edge ( 22 )
( 13 ) edge ( 32 );
\end{scope}
\end{tikzpicture}
{\scriptsize \textbf{(a)} GasLib-40 network.}
\end{minipage}
\hspace{0.03\linewidth}
\begin{minipage}[b]{.62\linewidth}
\footnotesize
\centering\begin{tikzpicture}[scale=.95]
\draw[thick, ->] (-.5,0) -- (6.5,0) node[right] {Inst.};
\draw[thick, ->] (0,-.5) -- (0,6) node[above] {Runtime in s};
\foreach \y in {2,4,6,8,10,12,14,16,18,20,22} { 
	\draw[thick] (.1, \y / 24 * 6.000 ) -- (-.1, \y / 24 * 6.000 ) node[left] {$\y$}; 
}

\draw[thick, dashed, black]
	(0,4.270 / 24 * 6.000) -- (6.25,4.270 / 24 * 6.000);
\draw[thick, dashdotted, black]
	(0,10.260 / 24 * 6.000) -- (6.25,10.260 / 24 * 6.000);

\foreach \point in {(0.120, 0.965), (0.240, 1.050), (0.360, 0.908), (0.480, 0.908), (0.600, 0.916), (0.720, 0.928), (0.840, 0.894), (0.960, 0.969), (1.080, 0.896), (1.200, 0.970), (1.320, 0.849), (1.440, 1.356), (1.560, 1.409), (1.680, 0.892), (1.800, 1.062), (1.920, 1.102), (2.040, 0.935), (2.160, 0.924), (2.280, 0.953), (2.400, 0.972), (2.520, 1.226), (2.640, 1.059), (2.760, 1.062), (2.880, 1.405), (3.000, 1.133), (3.120, 0.940), (3.240, 1.052), (3.360, 1.119), (3.480, 0.892), (3.600, 1.021), (3.720, 1.357), (3.840, 1.452), (3.960, 1.320), (4.080, 1.088), (4.200, 1.042), (4.320, 1.169), (4.440, 0.942), (4.560, 0.913), (4.680, 0.931), (4.800, 1.582), (4.920, 1.116), (5.040, 1.437), (5.160, 0.971), (5.280, 0.985), (5.400, 1.023), (5.520, 1.048), (5.640, 1.152), (5.760, 0.973), (5.880, 0.920), (6.000, 1.187)} {
	\fill[black] \point circle(2pt);
}
\foreach \point in {(0.120, 1.576), (0.240, 2.307), (0.360, 1.363), (0.480, 2.650), (0.600, 1.900), (0.720, 1.652), (0.840, 2.517), (0.960, 1.481), (1.080, 1.658), (1.200, 1.602), (1.320, 0.774), (1.440, 4.786), (1.560, 4.499), (1.680, 1.094), (1.800, 2.330), (1.920, 2.683), (2.040, 1.094), (2.160, 1.794), (2.280, 1.938), (2.400, 1.542), (2.520, 3.728), (2.640, 2.063), (2.760, 2.682), (2.880, 4.851), (3.000, 2.695), (3.120, 1.612), (3.240, 1.603), (3.360, 2.440), (3.480, 1.620), (3.600, 1.590), (3.720, 4.723), (3.840, 4.603), (3.960, 4.680), (4.080, 2.690), (4.200, 2.443), (4.320, 2.739), (4.440, 1.976), (4.560, 4.554), (4.680, 1.412), (4.800, 4.489), (4.920, 2.636), (5.040, 3.902), (5.160, 1.603), (5.280, 1.763), (5.400, 1.758), (5.520, 2.339), (5.640, 4.697), (5.760, 1.899), (5.880, 1.595), (6.000, 5.626)} {
	\fill[white, draw=black] \point circle(2pt);
}
\begin{scope}[shift={(-.55,-1.25)}]
\draw (-.45,.5) rectangle (8.1, -.5);
\fill[black] (0,.25) circle(2pt) node[right=.25cm, black] {MCA runtime};
\draw[thick, dashed, black]    (-.25,-.25) -- (.25, -.25) node[right, black] (xyz) {avg. MCA time {\tiny($4.3\,$s)}};
\fill[white, draw=black] (4.05,.25) circle(2pt) node[right=.25cm, black] {MCFI runtime};
\draw[thick, dashdotted, black]
(3.8,-.25) -- (4.3, -.25) node[right, black] {avg. MCFI time {\tiny($10.3\,$s)}};
\end{scope}
\end{tikzpicture} \\
{\scriptsize \textbf{(b)} Runtimes of both algorithms for random instances.}
\end{minipage}
\begin{minipage}[b]{.95\linewidth}
    \footnotesize
    \centering
    \input{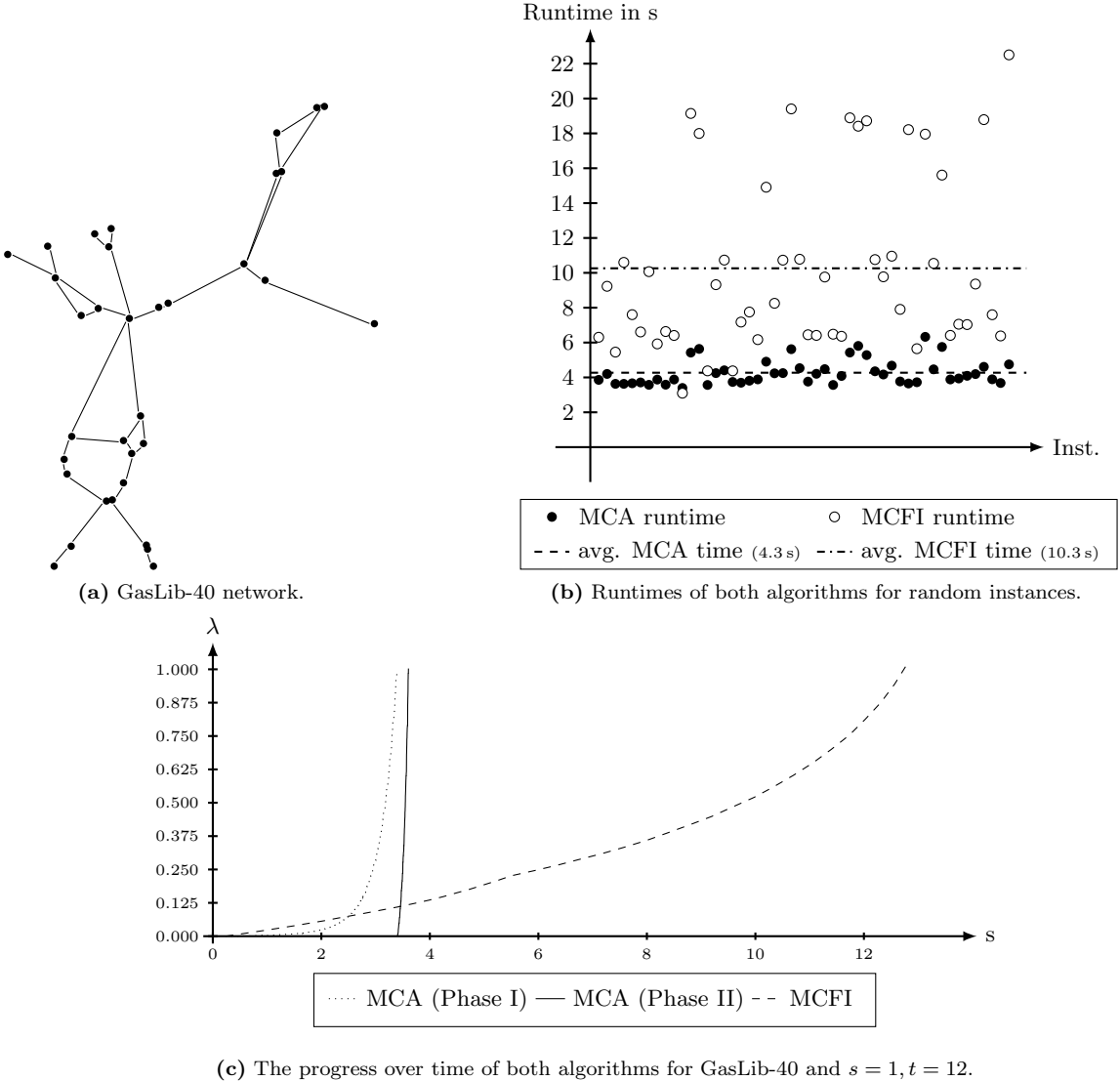}
    \\
    {\scriptsize \textbf{(c)} The progress over time of both algorithms for GasLib-40 and $s = 1, t=12$.}
\end{minipage}
\caption{Results of the computational study for the GasLib-40 gas network.}
\label{fig:gas40}
\end{figure}
%
%
%
%
%

As a specific example for the gas networks, we consider the GasLib-40 network that is depicted in Figure~\ref{fig:gas40}(a). This network is a small exemplary gas network that is not based on a real world example. The GasLib contains one scenario $\bar{\vec{b}}$ for in- and outflow rates at the vertices. We use the MCA algorithm and MCFI algorithm to compute parametric solutions for demands $\lambda \vec{b} +\bar{\vec{b}}$ with $\lambda \in [0, 1]$, where the demand vector $\vec{b}$ models an $s$-$t$-flow of rate $r = \frac{1}{2} \sum_{v \in V} |\bar{b}_{v}| = 2175$. We repeat the parametric computation $20$ times for random $s$-$t$-pairs and obtain an average runtime of $4.3$s for the MCA algorithm and $10.3$s for the MCFI algorithm. 
Figure~\ref{fig:gas40}(b) depicts the runtimes of both algorithms for all random instances. We see that runtimes are similar for most instances. However, for some choices of $s$ and $t$ the MCFI algorithm has a significantly higher runtime. In these instances, the Frank-Wolfe algorithm used to compute the fixed parameter solutions has a higher runtime slowing the MCFI algorithm down.

Figure~\ref{fig:gas40}(c) shows the progress of both algorithms over time. We use two runs of the MCA algorithm. In the first run, we run the algorithm for the demand function $\vec{b}(\lambda) = \lambda \bar{\vec{b}}$ in order to compute an initial solution for the non-zero demand $\bar{\vec{b}}$. We refer to this first run as Phase~I and depict its progress as a dotted line in Figure~\ref{fig:gas40}(c). In the second run of the MCA algorithm that we refer to as Phase~II (depicted by the solid line in Figure~\ref{fig:gas40}(c)) the actual parametric problem is solved starting from the initial solution obtained in the first phase. In the particular run shown in Figure~\ref{fig:gas40}(c), the interpolation of the cost function took about $0.3$s, Phase~I needed $3.1$, and Phase~II only took $0.1s$. We see that most of the runtime in this example is needed for the initial Phase~I. This can be explained with the fact Phase~I begins with very small flows (as it begins with small demands). Since the marginal cost in the gas networks are homogeneous (i.e., $\mcost(0) = 0$), the interpolation requires a very fine mesh around $0$ in order to guarantee the $(\alpha, \beta)$-approximation. Therefore, the electrical flow algorithm solving Phase~I traverses many regions for very small demands, leading to many iterations for small demands, as it can also be seen in Figure~\ref{fig:siouxfalls2}(b).

\subsubsection{Overall results}

\begin{table}[tb]
\centering
\begin{minipage}{\textwidth}
\centering
\def\arraystretch{1.05}
\begin{tabular}{lrrrrr}
	\toprule
	Network & $n$ & $m$ & \parbox{1.0cm}{\centering MCA} & \parbox{1.35cm}{\centering MCFI} & \parbox{0.8cm}{\centering FW} \\
	\midrule
	SiouxFalls &
		$24$ & $76$ & $0.3\,$s & $48.5\,$s\phantom{$^1$} & $0.3\,$s\\
	Berlin (Tiergarten) &
		$321$ & $545$ & $5.1\,$s & $135.4\,$s\phantom{$^1$} & $2.2\,$s\\
	Anaheim &
		$416$ & $914$ & $2.2\,$s & $57.4\,$s\phantom{$^1$} & $0.8\,$s\\
	Berlin (M.-P.-F.-C.) &
		$843$ & $1376$ & $76.9\,$s & $854.8\,$s\phantom{$^1$} & $12.5\,$s\\
	Chicago-Sketch &
		$546$ & $2176$ & $76.9\,$s & $> 3600.0\,$s\footnote{None of the 20 instances finished within the time limit of one hour.} & $77.2\,$s\\
	\midrule
	GasLib-11 &
		$8$ & $8$ & $0.4\,$s & $0.9\,$s\phantom{$^1$} & $0.0\,$s\\
	GasLib-24 &
		$18$ & $19$ & $1.1\,$s & $1.6\,$s\phantom{$^1$} & $0.1\,$s\\
	GasLib-40 &
		$34$ & $39$ & $4.3\,$s & $10.3\,$s\phantom{$^1$} & $0.2\,$s\\
	GasLib-134 &
		$87$ & $86$ & $10.1\,$s & $8.3\,$s\phantom{$^1$} & $0.1\,$s\\
	GasLib-135 &
		$106$ & $141$ & $81.7\,$s & $228.2\,$s\phantom{$^1$} & $1.0\,$s\\
	GasLib-582 &
		$268$ & $278$ & $475.5\,$s & $613.9\,$s\phantom{$^1$} & $0.7\,$s\\
	\bottomrule
\end{tabular}
\end{minipage}
\caption{The runtimes of the minimum-cost approximation (MCA), the minimum-cost flow interpolation (MCFI), and the Frank-Wolfe algorithm for a single fixed demand (FW) on several instances.}
\label{tab:studyresults}
\end{table}

Table~\ref{tab:studyresults} shows the results of both the MCA and the MCFI algorithms for all networks included in this computational study.
 For the traffic networks, the results are for $k$ randomly chosen source and sink pairs, with $k=50$ for the smaller networks (Sioux-Falls, Berlin-Tiergarten, Anaheim) and $k=20$ for the larger networks (Berlin Mitte-Prenzlauerberg-Friedrichshain-Center, Chigaco-Sketch). For the gas networks, a random scenario (if there was more than one possible scenario) is chosen, and we run the algorithm $k=20$ with an affine linear demand function based on the random scenario as offset and a random $s$-$t$-flow as demand direction. In addition to the MCA and MCFI algorithm, the table contains the runtime of the Frank-Wolfe (FW) algorithm for fixed parameter $\lambda = 1$ as a reference value. For all computations we use a time limit of one hour which was only exceeded for the largest traffic network, Chicago-Sketch.

For all networks, except for GasLib-134, the MCA algorithm is considerably faster than MCFI. The network GasLib-134 is an exception due to its structure. The network is a tree which makes the solution with the Frank-Wolfe subroutine of the MCFI algorithm extremely fast. Since the MCA method always needs some overhead time to compute the interpolation of the marginal cost functions, MCFI is faster in these examples.
We also note, that for the gas instances, the MCA times also include the time for Phase~I. If the initial solution is computed differently (for example with a single call of the Frank-Wolfe algorithm), these runtimes may be even improved. 
Overall, we conclude that both the MCA and the MCFI algorithm are applicable in practice. In particular for large instances, the MCA algorithm has considerably better runtimes.


%
%

\bibliography{computational_paper_arxiv.bib}   

\end{document}